\documentclass[aps,pra,twocolumn,floatfix,eqsecnum]{revtex4-1}
\usepackage{graphicx}
\usepackage{bm}
\usepackage{amssymb,amsfonts,textcomp}
\usepackage{multirow}
\usepackage[hypertex,colorlinks=true,linkcolor=blue,citecolor=blue,urlcolor=blue]{hyperref}
\usepackage{color}
\usepackage{array}
\usepackage{hhline}
\usepackage[normalem]{ulem}
\usepackage{enumerate}
\usepackage{amsmath}    
\usepackage{verbatim}   
\usepackage{color}      
\usepackage{subfigure}  

\def\hm#1{\frac{\hbar^2}{#1m}}
\def\rvec{ {\bf r}}
\def\kvec{ {\bf k}}
\def\he#1{$^{#1}$He}

\def\Bra#1{\left\langle #1\right|}
\def\Ket#1{\left| #1\right\rangle}
\def\bra#1{\bigl\langle #1\bigr|}
\def\ket#1{\bigl| #1\bigr\rangle}
\def\pd{{\phantom{\dagger}}}
\def\I{{\rm i}} 
\def\half{\frac{1}{2}}

\def\perz#1{\alpha_{#1}^{\dagger}}
\def\pver#1{\alpha_{#1}^{\phantom{\dagger}}}       
\def\qerz#1{a_{#1}^{\dagger}}
\def\qver#1{a_{#1}^{\phantom{\dagger}}}            
\def\qupdown{{\uparrow\downarrow}}
\def\qupup{{\uparrow\uparrow}}
\def\EF{e_{\rm F}}
\def\KF{k_{\rm F}}
\def\SF{S_{\rm F}}
\def\a0{a_0}

\begin{document}
\title{Correlations in the low-density Fermi gas:\\ Fermi-Liquid
state, Dimerization, and BCS Pairing} \author{H. H. Fan$^1$,
E. Krotscheck$^{1,2}$,
T. Lichtenegger$^{1,2}$, D. Mateo$^{2,3}$ and R. E. Zillich$^{2}$}
\affiliation{$^1$Department of Physics, University at Buffalo SUNY
Buffalo NY 14260}
\affiliation{$^2$Institut f\"ur Theoretische Physik, Johannes Kepler
Universit\"at, A 4040 Linz, Austria}
\affiliation{$^3$Dept. E.C.M., Facultat de Fisica, Universitat de Barcelona,
Spain}

\begin{abstract}

We present ground state calculations for low-density Fermi gases
described by two model interactions, an attractive square-well
potential and a Lennard-Jones potential, of varying strength. We use
the optimized Fermi-Hypernetted Chain integral equation method which
has been proved to provide, in the density regimes of interest here,
an accuracy better than one percent. We first examine the low-density
expansion of the energy and compare with the exact answer by Huang and
Yang (H. Huang and C. N. Yang, {\em Phys. Rev.\/} {\bf 105}, 767
(1957)). It is shown that a locally correlated wave function of the
Jastrow-Feenberg type does not recover the quadratic term in the
expansion of the energy in powers of $\a0\KF$, where $\a0$ is the
vacuum $s$-wave scattering length and $\KF$ the Fermi wave number. The
problem is cured by adding second-order perturbation corrections in a
correlated basis.  Going to higher densities and/or more strongly
coupled systems, we encounter an instability of the normal state of
the system which is characterized by a divergence of the {\em
  in-medium\/} scattering length. We interpret this divergence as a
phonon-exchange driven dimerization of the system, similar to what one
has at zero density when the vacuum scattering length $\a0$ diverges.
We then study, in the stable regime, the superfluid gap and its
dependence on the density and the interaction strength. We identify
two different corrections to low-density expansions: One is medium
corrections to the pairing interaction, and the other one finite-range
corrections. We show that the most important finite-range corrections
are a direct manifestation of the many-body nature of the system.
\end{abstract}

\maketitle
\clearpage

\section{Introduction}

The study of ultracold quantum gases is interlinked with controlling
magnetic Fano-Feshbach resonances and thereby changing the effective
interparticle interaction by many orders of
magnitudes~\cite{inouyeNature98,courteillePRL98,%
  timmermansPhysRep99,giorginiRMP08,chinRMP10}.  This makes ultracold
Fermi gases a convenient tool to study the behavior of a degenerate
fermionic many-body system~\cite{oharaScience02} over a wide range of
interaction strengths, in particular fermionic
superfluidity~\cite{chinScience04,BartensteinPRL04,kinastPRL04,%
  Zwierlein06,Chen06}.  Changing the magnetic field across a resonance
makes it possible to continuously tune the gas from the
Bardeen-Cooper-Schrieffer (BCS) state of Cooper pairs to a
Bose-Einstein condensate (BEC) of weakly bound molecules.  After the
first observations of a molecular BEC of fermions
\cite{jochimScience03,greinerNature03,zwierleinPRL03a}, this so-called
BCS-BEC cross-over has been widely studied experimentally
\cite{BartensteinPRL04,regalPRL04,regalPRL05,kinastPRL04,bourdelPRL04,%
  greinerPRL05,partridgePRL05} and theoretically
\cite{hollandPRL01,timmermansPhysLettA01,linghamPRL14},
  see also reviews Refs.~\onlinecite{duinePhysRep04,chenPhysRep05}.
  The observation of quantized vortices on both sides of the BCS-BEC
  cross-over provided an unambiguous proof of superfluidity by
  fermionic pairing~\cite{zwierleinNature05}. Recent work investigated
  the effect of partial polarization of a two-component Fermi gas on
  the Fermi liquid parameters \cite{navonScience10}, the nature of the
  transition from a BCS state to a state of a molecular
  BEC~\cite{huZwierleinScience12}, and quantifying the superfluid
  fraction in a Fermi gas by means of second
  sound~\cite{sidorenkovNature13}.

We study in this work the low-density properties of homogeneous Fermi
gases at zero temperature.  We use for our study a square-well and a
Lennard-Jones interaction potential.  Changing the interaction
strength (coupling constant) of the respective potential changes the
scattering length for two-body scattering $\a0$, which we refer to as
vacuum scattering length. The {\em in-medium scattering\/} length $a$
will be introduced later.  When $\a0\,<\,0$, {\em i.e.\/} the
interaction is effectively attractive, one expects BCS type pairing of
particles with opposite spin.  As $\a0\rightarrow -\infty$, a low
energy resonance of the two-body problem generates bound dimers.
$\a0\rightarrow -\infty$ is called the unitary limit, since the only
relevant length scale is the inverse Fermi wave number $\KF^{-1}$.
Increasing the attraction further, the Cooper pairs become bound
molecules and the fermionic nature of their constituents becomes less
visible.

We are in particular interested in structural quantities such as the
energetics, distribution functions, the stability of the system
against spinodal decomposition, dimerization, and BCS pairing.  We
utilize a quantitative method of microscopic many-body theory to
determine correlation effects, {\em i.e.\/} effects beyond the weak
coupling or mean-field approximations \cite{Landau5} that are
routinely applied at low densities.

In the low density limit, many quantities like, for example, the
ground state energy, depend only on the dimensionless parameter
$\KF\,\a0$~\cite{HuangYang57,Landau5}.  We are interested here in the
parameter range where this ``universal'' behavior ceases to persist
due to correlation effects. An example for correlation effects is the
pair distribution function, $g(r)$. In mean--field approaches, $g(r)$
is equal to the distribution function of the non-interacting Fermi
gas, $g_{\rm F}(r)$.  As we will see below, $g(r)$ deviates, in
particular for spin-antiparallel particles, substantially from $g_{\rm
  F}(r)$ when the absolute value of the scattering length becomes
large compared to the characteristic length $\sigma$ of the
interaction potential.

In the limit of weak attractive interactions, the system can be
described by a BCS type wave function. When the weak-coupling
approximation does not apply (for example for Lennard-Jones type
interactions), the pairing gap can be obtained by an extension of the
Jastrow-Feenberg variational method. The correlated BCS (CBCS) method
\cite{CBFPairing,KroTrieste} is reviewed in section~\ref{sec:CBCS},
see also Refs. \onlinecite{YangClarkBCS,HNCBCS,Fabrocinipairing} for a
similar implementation of the same ideas. The CBCS theory takes into
account {\em short-ranged\/} correlations analogously to the theory
for normal systems, and supplements these by the typical BCS
correlations. In its essence, CBCS provides a recipe for calculating
an effective interaction that enters the standard BCS formalism. An
alternative way to deal with the problem is a full Fermi Hypernetted
Chain (FHNC) summation for large gap parameters has been suggested by
Fantoni \cite{Fantonipairing,Fabrocinipairing2}, we will comment on this
approach further below.

Our paper is organized as follows: In Sec.~\ref{sec:ManyBodyTheory} we
will review briefly the basics of the correlated basis functions (CBF)
method. We call this approach ``generic'' many-body theory because the
same equations can be derived from Green's functions approaches
\cite{parquet1}, from Coupled Cluster theory \cite{BishopValencia} and
from an extension of density functional theory which includes pair
correlations. We evaluate in section \ref{ssec:lowdens} the
low-density limit and show that the exact formula \cite{HuangYang57}
is not reproduced by the Jastrow-Feenberg and/or the ``fixed node''
approximation. To correct this problem, we apply in section
\ref{ssec:CBF} perturbation theory in a correlated basis. We show that
second-order CBF corrections must be added to obtain the correct
low--density expansion.

In Sec. \ref{sec:CBCS} we will review the CBCS theory.  We will show
that the theory can be formulated in exactly the same way as ordinary
BCS theory. CBCS simply provides a prescription for deriving weak,
effective interactions from a strong, bare interaction.  Upon closer
inspection, the mapping of the bare interaction to an effective
interaction is closely related to the transition from the bare
interaction to the $T$-matrix used in the low-density expansion of BCS
theory \cite{Gorkov,PethickSmith,heiselbergPRL00}.

In Sec.~\ref{sec:results}, we present our results for the energy, the
pair distribution function, the in-medium scattering length, and the
gap energy as a function of Fermi wave number $\KF$ and the vacuum
scattering length $\a0$. The dynamical correlations can be
characterized by three regimes: For short distances, $r\approx
\sigma$, these correlations are, of course, determined by the
interaction. The intermediate regime is dominated by two-body
scattering where correlations decay as $1/r$.  A third, asymptotic
regime for $r$ larger than the average interparticle distance is
dominated by many-body effects where the correlations decay as
$1/r^2$.

We find an instability of the FHNC-Euler-Lagrange (FHNC-EL) solutions
accompanied by a divergence of the in-medium scattering length $a$.
This instability occurs well before the divergence of the vacuum
scattering length $\a0$; it is caused by the induced interaction
mediated by phonon exchange.  Thus, strongly bound dimers can be
formed at finite density even if the bare potential does not have a
bound state. Note that this is a property of the normal system, wave
functions of the type normally used only in the ``BEC'' regime
\cite{carlsonPRL03,astraPRL04,astraPRL05BCSBEC} would here be more
appropriate. We will return to this point further below.

In the regime $\a0 <0$ we solve the CBCS gap equation and show that
deviations from the simple BCS approximation can be separated into two
contributions. One stems from the density dependence of the in-medium
scattering length, the other one from the non-negligible momentum
dependence of the pairing interaction in the CBCS gap equation.

\section{Generic Many-Body Theory}
\label{sec:ManyBodyTheory}
\subsection{Variational wave functions}
\label{ssec:FHNC}

We start our discussion with the Jastrow-Feenberg theory for
a strongly interacting, translationally invariant {\em normal\/} system.
As usual, we assume a non-relativistic many-body Hamiltonian
\begin{equation}
H = -\sum_{i}\frac{\hbar^2}{2m}\nabla_i^2 + \sum_{i<j}
v(\rvec_i-\rvec_j)
\label{eq:Hamiltonian}
\end{equation}
where $v(r)$ is a  phenomenological, local interaction.
The method starts with a variational {\em
ansatz\/} for the wave function \cite{FeenbergBook}
\begin{align}
\Psi_0({\bf r}_1,\ldots,{\bf r}_N) &= \frac{1}{\cal N}
	F({\bf r}_1,\ldots,{\bf r}_N)
	\Phi_0({\bf r}_1,\ldots,{\bf r}_N)\label{eq:wavefunction}\\
F({\bf r}_1,\ldots,{\bf r}_N) &= \exp\frac{1}{2}
\left[\sum_{i<j}  u_2({\bf r}_i,{\bf r}_j)
+ \ldots\right]\label{eq:Jastrow}\\
{\cal N} &= \left\langle \Phi_0\right|
F^\dagger F \left|\Phi_0\right\rangle^{\frac{1}{2}}\,.
\end{align}
$\Phi_0({\bf r}_1,\ldots,{\bf r}_N)$ is a model state, normally a
Slater-determinant for fermions and $\Phi_0({\bf r}_1,\ldots,{\bf
  r}_N)=1$ for bosons, and $F$ is the correlation operator written in
the form (\ref{eq:Jastrow}). There are basically two ways to deal with
this type of wave function. In quantum Monte Carlo studies, the wave
function (\ref{eq:wavefunction}) is referred to as ``fixed node
approximation''. An optimal correlation function $F({\bf
  r}_1,\ldots,{\bf r}_N)$ is obtained by stochastic means
\cite{carlsonPRL03,changPRA04,astraPRL04,astraPRL05BCSBEC,%
  changPRL05,loboPRL06,burovskiPRL06,akkineniPRB07,morrisPRA10,%
  bulgacPRL06,liPRA11}. A decomposition into $n$-body correlations
$u_n({\bf r}_1,\ldots,{\bf r}_n)$ is then, of course, not necessary.
Alternatively, one can use diagrammatic methods, specifically the
optimized Fermi-hypernetted chain method for the calculation of
physically interesting quantities.  These diagrammatic methods have
been successfully applied to such highly correlated Fermi systems as
$^4$He and $^3$He at $T=0$~\cite{polish}. They are naturally expected
to work much better in the low density systems of interest here.  In
fact, we have shown in recent work \cite{ljium} that even the simplest
version of the FHNC-EL theory is accurate within better than one
percent at densities less than 25 percent of the ground state density
of liquid \he3.

Diffusion Monte Carlo calculations typically use a parametrized
Jastrow-Feenberg (JF) ansatz for importance sampling, where the
parameters are optimized by variational Monte Carlo calculations. JF
theory makes explicit use of the form (\ref{eq:Jastrow}).  It has been
shown \cite{parquet1} that triplet correlations contribute to the
ground state energy only in fourth order of the interactions. Even in
strongly interacting quantum fluids like the helium liquids, triplet
correlations contribute no more than five to ten percent to the ground
state energy \cite{PPA2,polish} in both isotopes. They are completely
negligible below approximately 25 percent of the respective
equilibrium densities. We can identify, at the low densities we are
concerned with here, the Jastrow-Feenberg approximation with the
fixed-node approximation in quantum Monte Carlo calculations.

The correlations $u_n({\bf r}_1,\ldots,{\bf r}_n)$ are obtained by minimizing
the energy, {\em  i.e.\/} by solving the Euler-Lagrange (EL) equations
\begin{eqnarray}
&&E_0 = \left\langle\Psi_0\right|H\left|\Psi_0\right\rangle
\equiv H_{{\bf o},{\bf o}}\label{eq:energy}\\
&&        \frac{\delta E_0}
{\delta u_n}({\bf r}_1,\ldots,{\bf r}_n) = 0\,.
\label{eq:euler}
\end{eqnarray}

The evaluation of the energy (\ref{eq:energy}) for the variational
wave function (\ref{eq:Jastrow}) and the analysis of the variational
problem are carried out by cluster expansion and resummation methods.
The procedure has been described at length in review articles
\cite{Johnreview,polish} and pedagogical material \cite{KroTrieste}.
In any approximate evaluation of the energy expectation value, it is
important to make sure that the resulting equations are consistent
with the {\em exact\/} variational determination of the
correlations. It has turned out that the (Fermi-)hypernetted chain
hierarchy of approximations is the only systematic approximation
scheme that preserves the properties of the exact variational
problem~\cite{FeenbergBook}.

Here, we spell out the simplest version of the equations that is
consistent with the variational problem (``FHNC-EL//0''). These do not
provide the quantitatively best implementation \cite{polish}. Instead,
they provide the {\em minimal\/} version of the FHNC-EL theory. In
particular, they contain the relevant physics, namely the correct
description of both short- and long--ranged correlations. They also
are the minimal implementation of the theory that gives the correct
expansion of the ground state energy in powers of $(\KF\,\a0)$ for the
wave function Eq. (\ref{eq:wavefunction}).

In the FHNC-EL//0 approximation \cite{Mistig}, which contains both the
random phase approximation (RPA) and the Bethe-Goldstone equation in a
``collective'' approximation \cite{KroTrieste}, the Euler equation
(\ref{eq:euler}) can be written in the form
\begin{equation}
S(k) = \frac{\SF(k)}{\sqrt{1 +
	2\frac{\displaystyle \SF^2(k)}{\displaystyle t(k)}
\tilde V_{\rm p-h}(k)}} \,,
\label{eq:FermiRPA0}
\end{equation}
where $S(k)$ is the static structure factor of the interacting system,
$t(k) = \hbar^2 k^2/2m$ is the kinetic energy of a free particle,
$\SF(k)$ is the static structure of the non-interacting Fermi system, 
and
\begin{eqnarray}
V_{\rm p-h}(r) &=&\>
\left[1+ \Gamma_{\!\rm dd}(r)\right]v(r)
+ \frac{\hbar^2}{m}\left|\nabla\sqrt{1+\Gamma_{\!\rm dd}(r)}\right|^2 \nonumber\\
&&+ \Gamma_{\!\rm dd}(r)w_{\rm I}(r)\nonumber\\
&\equiv& v_{\rm CW}(r) + \Gamma_{\!\rm dd}(r)w_{\rm I}(r)
\label{eq:VddFermi0}
\end{eqnarray}
is the so-called ``particle-hole interaction''. We have also above
defined the ``Clark-Westhaus effective interaction'' $v_{\rm CW}(r)$
\cite{Johnreview}.  As usual, we define the Fourier transform with a
density factor,
\begin{equation}
\tilde f(\kvec) \equiv \rho \int d^3 r e^{\I\kvec\cdot\rvec} f(\rvec)\,.
\label{eq:Fouri}
\end{equation}
Auxiliary quantities are the ``induced interaction''
\begin{equation}
\tilde w_{\rm I}(k)=-t(k)
\left[\frac{1}{\SF(k)}-\frac{1}{ S(k)}\right]^2
\left[\frac{S(k)}{\SF(k)}+\frac{1}{2}\right]
\label{eq:inducedFermi0}
\end{equation}
and the ``direct-direct correlation function'' 
\begin{equation}
\tilde \Gamma_{\!\rm dd}(k) = \bigl(S(k)-\SF(k)\bigr)/\SF^2(k)\,.
\label{eq:GFHNC}
\end{equation}
Eqs.~(\ref{eq:FermiRPA0})--(\ref{eq:GFHNC}) form a closed set which
can be solved by iteration.  Note that the Jastrow correlation
function (\ref{eq:Jastrow}) has been eliminated entirely.

\begin{figure}[h]
\centerline{\includegraphics[width=0.6\columnwidth]{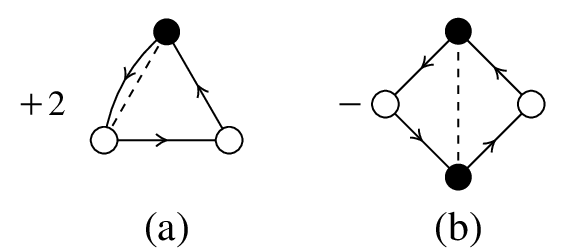}}
\caption{The two diagrams contributing to $(\Delta X_{\rm ee})_1(r)$.
The usual diagrammatic notations of FHNC-EL theory
\cite{KroTriesteBook} apply.}
\label{fig:Xee}
\end{figure}

The pair distribution function can generally be written as
\begin{equation}
g(r) = \left[1 + \Gamma_{\!\rm dd}(r)\right]\left[g_{\rm F}(r)+C(r)\right]\,.
\end{equation}
Roughly speaking, $\Gamma_{\!\rm dd}(r)$ describes dynamic,
short-range correlations, $g_{\rm F}(r) = 1-\frac{1}{\nu}\ell^2(r\KF)$
is the pair distribution function of the non-interacting Fermi gas,
and $\ell(x)=\frac{3}{x}j_1(x)$ is the Slater exchange
function. $C(r)$ describes the combination of statistical and dynamic
correlations.  In leading order in the dynamic correlations $
\Gamma_{\!\rm dd}(r)$ we have
\begin{equation}
\tilde C(k) = (S_{\rm F}^2(k)-1)\tilde\Gamma_{\!\rm dd}(k)
+ (\Delta \tilde X_{\rm ee})_1(k)
\label{eq:C0ofk}
\end{equation}
where $(\Delta \tilde X_{\rm ee})_1(k)$ is represented by the two
exchange diagrams shown in Fig. \ref{fig:Xee}.

In this approximation, the energy per particle is \cite{polish,ljium},
\begin{widetext}
\begin{eqnarray}
\frac{E}{N} &=&  \frac{3}{5}\EF  + e_{\rm R} +  e_{\rm Q} + t^{(3)}_{\rm JF}
\,,\nonumber \\
e_{\rm R} &=& \frac{\rho }{2}\int\! d^3r\>\bigl[g_{\rm F}(r) + C(r)\bigr]
v_{\rm CW}(r)\,,
\label{eq:EJF}\\
e_{\rm Q} &=& \frac{1}{4}\int\!\frac{d^3k}{(2\pi)^2\rho}\>
t(k)\tilde\Gamma_{\!\rm dd}^2(k)\left[S^2_{\rm F}(k)/S(k)-1\,\right]
-\frac{1}{4}\int\!\frac{d^3k}{(2\pi)^2\rho}\>
t(k)\tilde\Gamma_{\!\rm dd}(k)(\Delta \tilde X_{\rm ee})_1(k)
\equiv e_{\rm Q}^{(1)}+ e_{\rm Q}^{(2)}\label{eq:EQ}\\
t^{(3)}_{\rm JF} &=&  \frac{\hbar^2\rho^2}{8m\nu^2}
\int d^3 r_{12}d^3r_{13}
\Gamma_{\rm dcc}(\rvec_1;\rvec_2,\rvec_3)
\nabla_1^2\ell(r_{12}\KF)\ell(r_{13}\KF)
\label{eq:TJF}
\end{eqnarray}
where $\EF = \frac{\hbar^2\KF^2}{2m}$ is the Fermi energy of
non-interacting particles.  $\nu$ is the degree of degeneracy of the
single particle states; in our case we have generally $\nu=2$. The
term $t_{\rm JF}^{(3)}$ is the three-body term of the Jackson-Feenberg
kinetic energy.  The function $\Gamma_{\rm
  dcc}(\rvec_1;\rvec_2,\rvec_3)$ is the sum of all three-point
diagrams that have an exchange path connecting points $\rvec_2$ and
$\rvec_3$ and no exchange lines attached to point $\rvec_1$ which is
dynamically connected in such a way that there exists a path between
$\rvec_1$ and each of the other two external points that {\it does
  not\/} go trough the third external point.  The term $t^{(3)}_{\rm
  JF}$ is normally numerically very small; we must keep it here for
the purpose of deriving the low-density expansion.  To obtain the
correct low-density limit, we retain all contributions to $t_{\rm
  JF}^{(3)}$ with two factors $\Gamma_{\!\rm dd}(r)$:
\begin{eqnarray}
t^{(3)}_{\rm JF}&\approx& t_{\rm JF}^{(3a)}+ t_{\rm JF}^{(3b)}\nonumber\\
t_{\rm JF}^{(3a)} &=&  \frac{\hbar^2\rho^2}{8m\nu^2}
\int d^3 r_{12}d^3r_{13}\Gamma_{\!\rm dd}(r_{12})\Gamma_{\!\rm dd}(r_{13})
\nabla_{\rvec_1}^2\ell(r_{12}\KF)\ell(r_{13}\KF)\ell(r_{23}\KF)\\
t_{\rm JF}^{(3b)} &=& -\frac{\hbar^2\rho^2}{8m\nu^3}
\int d^3 r_{12}d^3r_{13}d^3r_{14}\Gamma_{\!\rm dd}(r_{13})
\Gamma_{\!\rm dd}(r_{24})
\nabla_{\rvec_1}^2\ell(r_{12}\KF)\ell(r_{13}\KF)\ell(r_{34}\KF)\ell(r_{24}\KF)\,.
\nonumber
\label{eq:TJF2}
\end{eqnarray}
The term $t_{\rm JF}^{(3b)}$ cancels exactly the contribution
to $e_{\rm Q}^{(2)}$ originating from the second diagram in
$(\Delta  X_{\rm ee})_1(r)$; the terms $e_{\rm Q}^{(2)}$, $t_{\rm JF}^{(3a)}$
and $t_{\rm JF}^{(3b)}$ can then be combined to
\begin{equation}
t_{\rm CW}^{(3)}= e_{\rm Q}^{(2)} + t_{\rm JF}^{(3a)} + t_{\rm JF}^{(3b)}
= \frac{\hbar^2\rho^2}{4m\nu^2}
\int d^3 r_{12}d^3r_{13}\nabla\Gamma_{\!\rm dd}(r_{12})\cdot\nabla
\Gamma_{\!\rm dd}(r_{13})\ell(r_{12}\KF)\ell(r_{13}\KF)\ell(r_{23}\KF)\,.
\label{eq:TCW}
\end{equation}
\end{widetext}
The term $t_{\rm CW}^{(3)}$ is recognized as the three-body
term of the ``Clark-Westhaus'' form of the kinetic energy.
To summarize, the total energy has the form
\begin{equation}
\frac{E}{N} = \frac{3}{5}\EF  + e_{\rm R} +  e_{\rm Q}^{(1)} + t_{\rm CW}^{(3)}\,.
\label{eq:Efull}
\end{equation}

For further reference, we also spell out the pair distribution functions
in the spin-parallel and the spin-antiparallel channel:
\begin{align}
g_{\qupup}(r) &=\left[1+\Gamma_{\!\rm dd}(r)\right]
\Big[1+\big[(S_{\rm F}^2(k)-1)\tilde\Gamma_{\!\rm dd}(k)\big]^{\cal F}(r)\nonumber\\
&\qquad\qquad\quad -\ell^2(r\KF )+2(\Delta X_{\rm ee})_1(r)\Big]\,,
\label{eq:guu}\\
g_{\qupdown}(r)  &=\left[1+\Gamma_{\!\rm dd}(r)\right]
\left[1+\big[(S_{\rm F}^2(k)-1)\tilde\Gamma_{\!\rm dd}(k)\big]^{\cal F}(r)
\right]
\label{eq:gud}\\
g(r) &= \frac{1}{2}\left[g_{\qupup}(r) + g_{\qupdown}(r)\right]
\label{eq:gfull}
\end{align}
where $\left[\ldots\right]^{\cal F}(r)$ indicates the
Fourier-transform (\ref{eq:Fouri}). To leading order in the density,
the term $\ell^2(r\KF )$ is the only term that reflects Fermi
statistics, whereas the factor $\left[1+\Gamma_{\!\rm dd}(r)\right]$
describes dynamical correlations.  $g_{\qupup}(r)$, $g_{\qupdown}(r)$, and $g(r)$
are normalized such that they go to unity for large $r$.

\subsection{Low-density limit}
\label{ssec:lowdens}

In the limit of low densities, the equation of state and related
quantities depend only on the vacuum $s$-wave scattering length $\a0$ and the
Fermi wave number $\KF$. For example, the energy per particle has the
expansion \cite{HuangYang57,Landau5}
\begin{equation}
\frac{E_{\rm HY}}{N} =
\frac{\hbar^2 \KF^2}{2m}\left[\frac{3}{5} + \frac{2}{3}\frac{\a0 \KF}{\pi} 
+ \frac{4(11-2\ln 2)}{35}\left(\frac{\a0 \KF}{\pi} \right)^2+ \ldots\right]
\label{eq:lowdensFermi}
\end{equation}
Note that the expansion (\ref{eq:lowdensFermi}) is strictly valid only
for $\a0 > 0$. For attractive potentials the superfluid condensation
energy must be added.

The locally correlated wave function (\ref{eq:wavefunction}) is not
exact, and the question arises whether it recovers the expansion
(\ref{eq:lowdensFermi}). It is plausible that this is {\em not\/} the
case: The calculation of the third term in Eq.
(\ref{eq:lowdensFermi}) makes explicit use of the form of the energy
denominator in second order perturbation theory \cite{Landau5}.  The
local correlation operator corresponds to a ``collective
approximation'' in which, among others, the particle-hole propagator
is approximated by a collective mode.

Our task is to express the variational energy expression
(\ref{eq:Efull}) to second order in the vacuum scattering length
$\a0$. One can deal with this task in two ways: One is to permit
hard-core interactions, the other, somewhat simpler, approach is to
assume a weak interaction that has a Fourier transform. In this case,
one can parallel the derivation of Ref. \onlinecite{parquet1} for
fermions.

We will show the details of the calculation in Appendix
\ref{app:lowdens}, here we discuss only the essential steps:
The vacuum scattering length is determined from the zero-energy
scattering equation
\begin{equation}
\frac{\hbar^2}{m}
\nabla^2\psi(r) = v(r)\psi(r)\,.
\label{eq:scatteq}
\end{equation}
The scattering equation has the asymptotic solution
\begin{equation}
\psi(r) = 1 - \frac{\a0}{r}\quad{\rm as}\quad r\rightarrow\infty .
\label{eq:a0def}
\end{equation}
Multiplying Eq. (\ref{eq:scatteq}) with $\psi(r)$ and using the identity
\begin{equation}
\psi(r)\nabla^2\psi(r) = \half\nabla^2 \psi^2(r) -
\left|\nabla \psi(r)\right|^2\,.
\label{eq:psi2}
\end{equation}
gives a relationship that will be useful later:
\begin{equation}
\frac{\hbar^2}{2m}
\nabla^2\psi^2(r) = \frac{\hbar^2}{m}
\left|\nabla \psi(r)\right|^2
+v(r)\psi^2(r)\equiv v_{\rm CW}^{(0)}(r)\,.
\label{eq:vcw}
\end{equation}
The quantity $v_{\rm CW}^{(0)}(r)$ is structurally identical to
$v_{\rm CW}$ as introduced in Eq.~(\ref{eq:VddFermi0}), except that $v_{\rm CW}^{(0)}(r)$
is calculated for the zero energy vacuum scattering solution $\psi(r)$.
Integrating Eq. (\ref{eq:vcw}) leads to the relationship
\begin{eqnarray}
\frac{4\pi\rho\hbar^2}{m}\a0 &=& \rho\int d^3r\left[\frac{\hbar^2}{m}
\left|\nabla \psi(r)\right|^2
+v(r)\psi^2(r)\right]\nonumber\\
&=& \tilde v_{\rm CW}^{(0)}(0+)\,.
\label{eq:afull}
\end{eqnarray}

We notice that the induced interaction $\tilde w_{\rm I}(k)$ as
defined in Eq. (\ref{eq:inducedFermi0}) is of second order in the
interaction. To leading order in the density we can also expand
Eq. (\ref{eq:FermiRPA0})
\[S(k) = \SF(k) - \frac{\SF^3(k)}{t(k)}\tilde v_{\rm CW}(k)\]
and obtain from Eq. (\ref{eq:GFHNC}) the solution
\begin{equation}
\tilde \Gamma_{\!\rm dd}(k) = -\frac{\tilde v_{\rm CW}(k)\SF(k)}{t(k)}\,.
\label{eq:EulerLow0}
\end{equation}

In addition to calculating the energy contributions (\ref{eq:Efull})
for the correlation function (\ref{eq:EulerLow0}), we must express
$\tilde v_{\rm CW}(0+)$ in terms of the scattering length because
$\tilde v_{\rm CW}(0+)$ is calculated with the optimal correlation
function (\ref{eq:GFHNC}) of the many-body problem at {\rm finite
density,} and not with the solution of the zero-density scattering
equation (\ref{eq:scatteq}). In Appendix \ref{app:lowdens} we will
prove the relationship

\begin{eqnarray}
\tilde v_{\rm CW}(0)&=& \frac{4\pi\rho\hbar^2}{m}\a0\nonumber\\
&+& \frac{1}{2} \int\frac{d^3 k}{(2\pi)^3\rho}\frac{\tilde v_{\rm CW}^2(k)}{t(k)}
\left[\SF(k) -1\right]^2 + {\cal O}(a_0^2)
\nonumber\\
&=&\frac{4\pi\rho\hbar^2}{m}\a0
\left[1+\frac{99}{280}\frac{\tilde v_{\rm CW}(0+)}{\EF}\right]\nonumber\\
&=& \frac{4\pi\rho\hbar^2}{m}\a0\left[1+\frac{33}{35}\frac{\a0\KF}{\pi}\right]\,.
\label{eq:vcw0}
\end{eqnarray}

Collecting all results, one finds
\begin{equation}
\frac{E}{N} =\frac{\hbar^2 \KF ^2}{2m}\left[\frac{3}{5}
+ \frac{2}{3}\frac{ a_0 \KF }{\pi} +  1.5415
\left(\frac{a_0 \KF }{\pi}\right)^2 +\ldots\right]\,,
\label{eq:lowdensfhnc0}
\end{equation}
see Appendix \ref{app:lowdens} for details of the calculation.  The
result (\ref{eq:lowdensfhnc0}) is to be compared with the factor
$4(11-2\ln 2)/35 = 1.098$ of Eq. (\ref{eq:lowdensFermi}). To get the
exact result, one must go beyond local correlation operators; this is
done by perturbation theory in a correlated basis generated by the
correlation operator $F(\rvec_1,\ldots,\rvec_N)$ described in the next
section \ref{ssec:CBF}.  The situation is analogous to the case of the
high-density limit of the correlation energy of the electron gas. With
local correlations one obtains for the logarithmic term 0.05690$\,\ln
r_s\,$Ry \cite{Zab80} instead of the exact value 0.06218$\,\ln
r_s\,$Ry \cite{Mac50,GellMannBrueckner}. This deficiency is, for the
electron gas, removed by second-order CBF theory
\cite{LanttoKroSmithOaxtepec}.  One conclusion of our analysis is that
the fixed node approximation in conjunction with Green's functions
\cite{changPRA04} or Diffusion \cite{astraPRL04,astraPRL05BCSBEC}
should reproduce the expansion (\ref{eq:lowdensfhnc0}) and not
(\ref{eq:lowdensFermi}).

\subsection{Elements of Correlated Basis Functions}
\label{ssec:CBF}

We have seen above that a locally correlated wave function
(\ref{eq:wavefunction}) does not produce the exact low-density limit
(\ref{eq:lowdensFermi}) of the ground state energy. As mentioned
above, the problem can be cured by applying second-order perturbation
theory with correlated basis functions (CBF theory). We will also need
the basic ingredients of CBF theory for examining the superfluid
state. We review the method only very briefly, details may be found in
pedagogical material \cite{KroTrieste} and review articles
\cite{Johnreview,polish}. The diagrammatic construction of the
relevant ingredients has been derived in Ref. \onlinecite{CBF2}.

CBF theory uses the correlation operator
$F$ to generate a complete set of correlated and normalized
$N$-particle basis states through
\begin{equation}
\vert \Psi_{\bf m}^{(N)} \rangle =
\frac{F_{\!N} \; \vert \Phi_{\bf m}^{(N)} \rangle }
{\langle \Phi_{\bf m}^{(N)} \vert  F_{\!N}^{\dagger} F^{\phantom{\dagger}}_{\!N}
\vert \Phi_{\bf m}^{(N)}
\rangle^{1/2} } \;,
\label{eq:States}
\end{equation}
where the $\{\vert \Phi_{\bf m}^{(N)} \rangle\}$ form a complete basis of
model states, normally consisting of Slater determinants of single
particle orbitals.  Although the $\vert \Psi_{\bf m}^{(N)} \rangle$ are not
orthogonal, perturbation theory can be formulated in terms of these
states \cite{MF1,FeenbergBook}.

For economy of notation, we introduce a ``second--quantized''
formulation of the correlated states. The Jastrow--Feenberg
correlation operator in (\ref{eq:Jastrow}) depends on the particle
number, {\it i.e.\/} $F=F_{\!N}(1,\ldots,N)$
(whenever unambiguous, we omit the corresponding subscript). Starting
from the conventional $\qerz{k}, \qver{k}$ that create and annihilate
single particle states, new creation and annihilation operators
$\perz{k},\pver{k}$ of {\em correlated states\/} are defined by their
action on the correlated basis states:
\begin{eqnarray}
\perz{k}\,\bigl|\Psi_{\bf m}\bigr\rangle
&\equiv\>& \frac{ F_{\!\!_{N+1}} \qerz{k} \,\ket {\Phi_{\bf m}} }{
\bra {\Phi_{\bf m}} \qver{k} F_{\!\!_{N+1}}^\dagger 
 F_{\!\!_{N+1}}^{\phantom{\dagger}}
 \qerz{k}\ket {\Phi_{\bf m}}^{1/2} }\, ,
\label{eq:creation}\\
\pver{k}\,\bigl|\Psi_{\bf m}\bigr\rangle
&\equiv\>& \frac{ F_{\!\!_{N-1}} \qver{k}\,\ket {\Phi_{\bf m}} }{
\bra {\Phi_{\bf m}} \qerz{k} F_{\!\!_{N-1}}^\dagger 
F_{\!\!_{N-1}}^{\phantom{\dagger}}
a_{k}\ket {\Phi_{\bf m}}^{1/2} }\,.
\label{eq:annihilation}\end{eqnarray}
According to these definitions, $\alpha_{k}^\dagger$ and
$\alpha^{\phantom{\dagger}}_{k}$ obey the same commutation rules as
the creation and annihilation operators $\qerz{k}$ and $\qver{k}$ of
uncorrelated states, {\it but they are not Hermitian conjugates.\/} If
$\ket{\Psi_{\bf m}}$ is an $N$--particle state, then the state in
Eq.~(\ref{eq:creation}) must carry an $(N\!+\!1)$-particle correlation
operator $F_{\!\!_{N+1}}$, while that in Eq.~(\ref{eq:annihilation})
must be formed with an $(N\!-\!1)$--particle correlation operator
$F_{\!\!_{N-1}}$.

In general, we label ``hole'' states which are occupied in $\vert
\Phi_{\bf o} \rangle$ by $h$, $h'$, $h_i\;, \ldots\,$, and unoccupied
``particle'' states by $p$, $p'$, $p_i\;,$ \textit{etc.}.  To display
the particle-hole pairs explicitly, we will alternatively to the
notation $\bigl|\Psi_{\bf m}\bigr\rangle$ use
$\bigl|\Psi_{p_1 \ldots p_d\, h_1\ldots h_d}\bigr\rangle $.  A basis
state with $d$ particle-hole pairs is then
\begin{equation}
\bigl|\Psi_{p_1 \ldots p_d\, h_1\ldots h_d}\bigr\rangle 
=\perz{p_1}\cdots\perz{p_d}\pver{h_d}\cdots\pver{h_1}\ket{\Psi_{\bf o}}
\,.
\label{eq:psimph}
\end{equation}

For the off--diagonal elements $O_{\bf m,n}$ of an operator $O$, we
sort the quantum numbers $m_i$ and $n_i$ such that $|\Psi_{\bf m}
\rangle$ is mapped onto $\left|\Psi_{\bf n}\right\rangle$ by
\begin{equation}
\label{eq:defwave}
\left|\Psi_{\bf m}\right\rangle = \perz{m_1}\perz{m_2}
\cdots 
\perz{m_d} \; \pver{n_d} \cdots \pver{n_2}\pver{n_1}  
\left|\Psi_{\bf n} \right\rangle \;.
\end{equation}
From this we recognize that, to leading order in $N$, any $O_{\bf
m,n}$ depends only on the {\it difference\/} between the states
$\vert \Psi_{\bf m} \rangle$ and $\vert \Psi_{\bf n} \rangle$, and {\it
not\/} on the states as a whole.  Consequently, $O_{\bf m,n}$ can be
written as matrix element of a $d$-body operator
\begin{equation}
\label{eq:defmatrix}
O_{\bf m,n} \equiv \langle m_1\, m_2 \, \ldots m_d \,| 
{\cal O}(1,2,\ldots d) \,|n_1\,
n_2 \, \ldots n_d\rangle_a \;.
\end{equation}
(The index $a$ indicates antisymmetrization.) 

The key quantities for the execution of the theory are diagonal and
off-diagonal matrix elements of unity and $H'\!\equiv H\!-\!H_{{\bf
    o},{\bf o}}$,
\begin{eqnarray}
M_{\bf m,n} &=& \langle \Psi_{\bf m} \vert \Psi_{\bf n} \rangle
\equiv \delta_{\bf m,n} +  N_{\bf m,n}\;,
\label{eq:defineNM}
\\
H'_{\bf m,n} &\equiv &
W_{\bf m,n} + \frac{1}{2}\left(H_{\bf m,m}+H_{\bf n,n}-2H_{\bf o,o}
\right)N_{\bf m,n} \,. \qquad
\label{eq:defineW}
\end{eqnarray}
Eq. (\ref{eq:defineW}) defines a natural decomposition
\cite{CBF2,KroTrieste} of the matrix elements of $H'_{\bf m,n}$ into
the off-diagonal quantities $W_{\bf m,n}$ and $N_{\bf m,n}$ and
diagonal quantities $H_{\bf m,m}$.

To leading order in the particle number, the {\it diagonal\/} matrix
elements of $H'\!\equiv H\!-\!H_{{\bf o},{\bf o}}$ become additive, so
that for the above $d$-pair state we can define the CBF single
particle energies
\begin{equation}
\bra{\Psi_{\bf m}} H' \ket{\Psi_{\bf m}} \>\equiv\>
\sum_{i=1}^d e_{p_ih_i}  + {\cal O}(N^{-1}) \;,
\label{eq:CBFph}
\end{equation}
with $e_{ph} = e_p - e_h$ where
\begin{eqnarray}
e_p &=&\phantom{-}\bigl\langle\Psi_{\bf o}\bigr|\pver{p}\,H'\perz{p}\
\bigl|\Psi_{\bf o}\bigr\rangle = t(p) + u(p)\nonumber\\
e_h &=&-\bigl\langle\Psi_{\bf o}\bigr|\perz{h}\,H'\pver{h}\
\bigl|\Psi_{\bf o}\bigr\rangle = t(h) + u(h)\,
\label{eq:spectrum}
\end{eqnarray}
and $u(p)$ is an average field that can be expressed in terms of the
compound diagrammatic quantities of FHNC theory.
According to (\ref{eq:defmatrix}),
$W_{{\bf m},{\bf n}}$  and $N_{{\bf m},{\bf n}}$ define 
$d-$particle operators ${\cal N}$ and ${\cal W}$, {\em e.g.\/}
\begin{eqnarray}
N_{{\bf m},{\bf o}} &\equiv& N_{p_1p_2\ldots p_d \,h_1h_2\ldots h_d,0} \nonumber\\
&\equiv& \langle p_1p_2\ldots p_d \,|\, {\cal N}(1,2,\ldots,d)\,
|\,h_1h_2\ldots h_d \rangle_a  \;,\nonumber\\
W_{{\bf m},{\bf o}} &\equiv& W_{p_1p_2\ldots p_d \,h_1h_2\ldots h_d,0}\nonumber\\
&\equiv&  \langle p_1p_2\ldots p_d \,|\, {\cal W}(1,2,\ldots,d)\,
|\,h_1h_2\ldots h_d \rangle_a  \;.\qquad\;
\label{eq:NWop}
\end{eqnarray}
Diagrammatic representations of ${\cal N}(1,2,\ldots,d)$ and ${\cal
W}(1,2,\ldots,d)$ have the same topology \cite{CBF2}.  In
homogeneous systems, the continuous parts of the $p_i,h_i$ are wave
numbers ${\bf p}_i,{\bf h}_i$; we abbreviate their difference as ${\bf
q}_i$.

In principle, the ${\cal N}(1,2,\ldots,d)$ and ${\cal
W}(1,2,\ldots,d)$ are non-local $d$-body operators. In the next
section, we will show that we need, for examining pairing phenomena,
only the two-body operators. Moreover, the low density of the systems
we are examining permits the same simplifications of the FHNC theory
that we have spelled out in Sec. \ref{ssec:FHNC}. In the same
approximation, the operators ${\cal N}(1,2)$ and ${\cal
W}(1,2)$  are local, and we have \cite{polish}
\begin{eqnarray}
{\cal N}(1,2) &=& {\cal N}(r_{12}) = \Gamma_{\rm dd}(r_{12})\nonumber\\
{\cal W}(1,2) &=& {\cal W}(r_{12})\,,\quad \tilde {\cal W}(k) =
- \frac{t(k)}{\SF(k)}\tilde \Gamma_{\rm dd}(k)\,.
\label{eq:NWloc}
\end{eqnarray}

The most straightforward application of CBF theory is to calculate
corrections to the ground state energy. In second order we have,
for example,
\begin{widetext}
\begin{equation}
\delta E_2 = -\frac{1}{4}\sum_{pp'hh'}\frac{
\left|\left\langle pp'\right|{\cal W}\left|hh'\right\rangle_a
+ \frac{1}{2}\left[t_p + t_{p'}-t_h -t_{h'}\right]
\left\langle pp'\right|{\cal N}\left|hh'\right\rangle_a\right|^2}
{t_p + t_{p'}-t_h -t_{h'}}\,.
\label{eq:E2CBF}
\end{equation}
\end{widetext}
The magnitude of the CBF correction is normally comparable to the
correction from three-body correlations \cite{polish}. It is also
important to note that there are significant cancellations between the
two terms in the numerator. We will show in appendix \ref{app:lowdens}
that the CBF correction (\ref{eq:E2CBF}) corrects the coefficient of
the third term in the expansion (\ref{eq:lowdensfhnc0}) and leads to
the exact low-density limit (\ref{eq:lowdensFermi}).

\section{BCS Theory with correlated wave functions}
\subsection{General derivation}
\label{sec:CBCS}

We show in this section how the variational theory is generalized to a
superfluid or superconducting state. We restrict ourselves here to the
simplest case of $s$--wave pairing and show how the effective
interactions, which enter phenomenological theories as parameters, may
be calculated from first principles.  This section reviews the
derivations of Refs. \onlinecite{CBFPairing,KroTrieste}.

The BCS theory of fermion superfluidity ge\-ne\-ra\-li\-zes the
Hartree--Fock model $\left\{\Ket{\Phi_m}\right\} $ by introducing a
superposition of independent particle wave functions corresponding to
different particle numbers \cite{BeliaevLesHouches}
\begin{equation}
\ket{{\rm BCS}} = \prod_{\kvec}
(u_{\kvec} + v_{\kvec} a_{\kvec ,\uparrow }^\dagger
a_{-{\kvec,\downarrow }}^\dagger)\ket{0} \, .
\label{8.6.1}
\end{equation}
The coefficient functions $u_{\kvec}$ and $v_{\kvec}$ are known as
Bogoliubov amplitudes. They describe the distortion of the Fermi
surface due to the pairing phenomenon.

To deal with strongly interacting systems, adequate provision must be
made for the singular or near--singular nature of the two--body
interaction $v(r)$ for small interparticle distances $r$.  To build
the required geometrical correlations into the microscopic description
of the system, we can define a correlated BCS state, incorporating
both short-ranged and BCS correlations.  We are faced with a formal
mismatch, which prevents us from simply applying the correlation
factor $F_{\!N}$ to $\ket{\rm BCS}$. The former is defined in the
$N$--particle Hilbert space and the latter is a vector in Fock space
with indefinite particle number.  The most natural way to deal with
this is first projecting the bare BCS state on an arbitrary member of
a complete set of independent--particle states with fixed particle
numbers, applying the correlation operator to that state,
normalizing the result, and finally summing over all particle numbers.
We must therefore distinguish between correlation operators and
normalization integrals corresponding to different particle numbers
$N$.  Thus, the correlated BCS (CBCS) state is
\begin{eqnarray}
\ket{\rm CBCS} &=& \prod_{\kvec}
(u_{\kvec} + v_{\kvec} \alpha_{\kvec ,\uparrow }^\dagger
\alpha_{-{\kvec,\downarrow }}^\dagger)\ket{\Phi_0}\nonumber\\
&=&  \sum_{m,N} \ket {\Psi_m^{(N)}}
\langle\Phi_m^{(N)} \ket{\rm BCS} \,.
\label{8.6.25}
\end{eqnarray}
The trial state (\ref{8.6.25}) superposes the correlated basis states
$\ket {\Psi_m^{(N)}}$ with the same amplitudes with which the model
states $\ket{\Phi_m^{(N)}}$ enter the corresponding expansion of the
{\it original\/} BCS vector.

To derive the relevant equations we consider the expectation value of
an arbitrary operator $\hat O$ with respect to the
superfluid state:
\begin{equation}
\left\langle\hat O\right\rangle_s  =
	\frac{{\bra {\rm CBCS}} \hat O {\ket {\rm CBCS}}}
	{\langle {\rm CBCS}\ket {\rm CBCS}} \, .
\label{8.6.26}
\end{equation}

One may pursue cluster--expansion and resummation methods of
expectation values (\ref{8.6.26}) for the superfluid trial state
(\ref{8.6.25}).  This has been done successfully for the one-- and
two--body density matrices corresponding to a slightly different
choice of the correlated BCS state~\cite{Fantonipairing} which
exhibits, unfortunately, divergences for optimized correlation
functions. We do not follow this route, but instead consider the
interaction of only one Cooper pair at a time. The error introduced by
this is of order $\xi = (\Delta_F / \epsilon_F )^2$, where $\Delta_F$
is the superfluid gap energy. We will demonstrate below that this
quantity is indeed small in the regime where the wave function
(\ref{8.6.25}) is appropriate.

In leading order, it is sufficient to retain the terms of {\it first
  order in the deviation\/} $v^2_{\kvec} - v_{0,\kvec}^2$ and those of
{\it second order\/} in $u_{\kvec}v_{\kvec}$. We refer to this as the
``decoupling approximation''. The calculation of $\bigl\langle \hat H
- \mu \hat N\bigr\rangle$ for correlated states \cite{CBFPairing} is
somewhat tedious, we only give the essential steps and the final
result.  It is convenient to introduce the creators and annihilators
of correlated Cooper pairs,
\begin{eqnarray}
\beta_\kvec^\dagger  &=& 
\alpha_{ \kvec\uparrow }^\dagger\alpha_{ -\kvec\downarrow }^\dagger
\nonumber\\
\beta_\kvec  &=& 
\alpha_{ -\kvec\downarrow }\alpha_{\kvec \uparrow } \, .
\label{5.7.8}
\end{eqnarray}
\begin{widetext}
In terms of these quantities, the expectation value of
an operator $\hat O$ is, to leading order in the amplitudes
$v^2_{\kvec} - v_{0,\kvec}^2$ and $u_{\kvec}v_{\kvec}$
\begin{eqnarray}
\langle \hat O \rangle_s
&=&	\Bigl\langle \Psi_0 \Bigr| O^{(N)}\Bigl| \Psi_0 \Bigr\rangle
\nonumber\\
&+&	\sum_{ k > \KF  } v_\kvec^2
\Bigl\langle \Psi_0\, \beta_\kvec  \Bigr|
	\left[ \hat O^{(N+2)} - O_{oo}^{(N)}\right]
	\Bigl|\beta_\kvec^\dagger \Psi_0 \Bigr\rangle
+	\sum_{ k < \KF  } u_\kvec^2
\Bigl\langle \Psi_0 \, \beta_k^\dagger \Bigr|
	\left[ \hat O^{(N-2)} - O_{oo}^{(N)}\right]
	\Bigl|\beta_{\kvec} \Psi_0 \Bigr\rangle
\nonumber\\
&+&\sum_{ k>\KF  , k' < \KF  } u_\kvec v_\kvec u_{\kvec'} v_{\kvec'}
\Bigl\langle \Psi_0 \Bigr|\left[\hat O^{(N)} - O_{oo}^{(N)}\right]
	\Bigl|\beta_\kvec^\dagger \beta_{\kvec'}\Psi_0 \Bigr\rangle
\nonumber\\
&+&	\sum_{ k > \KF  , k' > \KF  } u_\kvec v_\kvec u_{\kvec'} v_{\kvec'}
\Bigl\langle \Psi_0 \beta_\kvec \Bigr|
	\left[\hat O^{(N+2)} - O_{oo}^{(N)}\right]
\Bigl|\beta_{\kvec'}^\dagger\Psi_0 \Bigr\rangle
\nonumber\\
&&+	\sum_{ k < \KF  , k' < \KF  } u_\kvec v_\kvec u_{\kvec'} v_{\kvec'}
\Bigl\langle \Psi_0 \beta_\kvec^\dagger\Bigr|
	\left[ \hat O^{(N-2)} - O_{oo}^{(N)}\right]
	 \Bigl|\beta_{\kvec'} \Psi_0 \Bigr\rangle
\nonumber\\
&+&
\sum_{ k < \KF  , k' > \KF  } u_\kvec v_\kvec u_{\kvec'} v_{\kvec'}
\Bigl\langle \Psi_0 \beta_\kvec^\dagger \beta_{\kvec'}\Bigr|
	\left[\hat O^{(N)} - O_{oo}^{(N)}\right]
	\Bigl| \Psi_0 \Bigr\rangle\,.
\label{5.7.9}
\end{eqnarray}
In Eqs. (\ref{5.7.9}), the operators $\hat O^{(N)}$, $\hat O^{(N-2)}$,
and $\hat O^{(N+2)}$ are the $N$, $N-2$, and $N+2$ --particle
realizations of the operator $\hat O$, and $O_{oo}^{(N)}$ the
expectation value of the operator $\hat O$ in the $N$--particle
correlated ground state.  Inserting $\hat H - \mu \hat N$ for $\hat O$
into the expansion (\ref{5.7.9}), where $\mu $ is a Lagrange
multiplier (the chemical potential) introduced to adjust the average
particle number $\langle\hat N \rangle_s = N$, we recover the
effective interactions, overlap integrals (\ref{eq:NWop}), and
single--particle energies (\ref{eq:spectrum}) of section
\ref{ssec:CBF}, {\it e.g.}
\begin{eqnarray}
\bra {\Psi_0\, \beta_\kvec}
\left[\hat H^{(N+2)} - \mu(N+2) - H_{oo}^{(N)} + \mu N\right]
\ket{ \beta_\kvec^\dagger\,\Psi_0} &=&	2 [ e_k- \mu ]\,,
\label{5.7.10}\\
\bra {\Psi_0}\left[\hat H^{(N)}-\mu\hat N - (H_{oo}^{(N)}-\mu N)\right]
	\ket{\beta_\kvec^\dagger \beta_{\kvec'}^\pd\, \Psi_0}
&=&\bra {\kvec\uparrow,-\kvec\downarrow} W (1,2)
	\ket {\kvec'\uparrow , -\kvec'\downarrow}_a
\label{5.7.11}\\
&+& \left[ e (k) - e (k') \right]
\bra {\kvec\uparrow, -\kvec\downarrow}
	N (1,2)\ket {\kvec'\uparrow, - \kvec'\downarrow}_a\,,
\nonumber\\
\bra {\Psi_0\beta_\kvec^\pd}\left[ \hat H^{N+2} -
\mu(N+2)- (H_{oo}^{(N)} - \mu N)\right]
\ket{\beta_{\kvec'}^\dagger\Psi_0}
& =& \bra {\kvec\uparrow, -\kvec\downarrow} W(1,2)
\ket{\kvec'\uparrow,-\kvec'\downarrow}_a
\label{5.7.12}\\
&+& (e(k) + e(k') - 2\mu)
\bra{\kvec\uparrow , -\kvec\downarrow} N(1,2)
	\ket{ \kvec'\uparrow,-\kvec'\downarrow}_a\,.\nonumber
\end{eqnarray}

Accordingly, we may write the energy of the superfluid state
in the form
\begin{equation}
\langle \hat H - \mu \hat N \rangle_s = H_{oo}^{(N)} - \mu
N + 2 \sum_{ k>\KF  } v_k^2 (e_k- \mu ) + 2
\sum_{ k<\KF  } u_k^2 (e_k- \mu ) + \sum_{\kvec,\kvec'}u_\kvec v_\kvec
u_\kvec' v_\kvec' {\cal P}_{\kvec\kvec'}
\label{5.7.13}\end{equation}
with the ``pairing interaction''
\begin{eqnarray}
{\cal P}_{\kvec\kvec'} &=& \bra{\kvec \uparrow ,-\kvec\downarrow}
{\cal W}(1,2)\ket{\kvec'\uparrow ,-\kvec'\downarrow}_a
+ (|e_k- \mu | + |e_{k'}- \mu |)
\bra{\kvec \uparrow ,-\kvec\downarrow}
{\cal N}(1,2)\ket{\kvec'\uparrow , - \kvec'\downarrow}_a\nonumber\\
&\equiv&{\cal W}_{\kvec\kvec'}+(|e_k- \mu | + |e_{k'}- \mu |)
{\cal N}_{\kvec\kvec'}\,.
\label{5.7.14}\end{eqnarray}
\end{widetext}
With the results (\ref{5.7.13}) and (\ref{5.7.14}), we have arrived at
a formulation of the theory which is formally identical to the BCS
theory for weakly interacting systems. Upon closer inspection (see the
next section) we will see that our formulation corresponds to a BCS
theory formulated in terms of the scattering matrix
\cite{PethickSmith}. The correlation operator serves here to tame the
short--range dynamical correlations. The effective interaction ${\cal
  W}(1,2)$ is just an energy independent approximation of the
$T$-matrix.

We may proceed now in the conventional way to determine the
Bogoliubov--amplitudes $u_\kvec $, $v_\kvec $, by variation of the
condensation energy (\ref{5.7.13}) to compute the superfluid
condensation energy or to investigate the local stability of the
normal ground state by second variation. Minimization of the energy
expectation value determines the BCS amplitudes $u_{\kvec}$,
$v_{\kvec}$.  The CBCS gap equation becomes
\begin{equation}
\Delta_\kvec = -\frac{1}{2}\sum_{\kvec'} {\cal P}_{\kvec\kvec'}
\frac{\Delta_{\kvec'}}{\sqrt{(e_{\kvec'}-\mu)^2 + \Delta_{\kvec'}^2}}\,.
\label{eq:gap}
\end{equation}
The conventional, {\em i.e.\/} uncorrelated, BCS gap equation
\cite{FetterWalecka} is retrieved by replacing the effective
interaction ${\cal P}_{\kvec\kvec'}$ by the matrix elements of the
bare interaction.  Note that our ``decoupling approximation'' simply
means that we assume that the pairing interaction ${\cal
P}_{\kvec\kvec'}$ does not depend on the Bogoliubov amplitudes.

To conclude this section, we point to a subtle issue concerning the
normalization of the correlated BCS state (\ref{8.6.25}).  Above, we
have written the wave function as a specific linear combination of
{\em normalized\/} states $\ket{\Phi_m^{(N)}}$. In our formulation
of a correlated pairing theory \cite{HNCBCS} as well as in related work
\cite{Fantonipairing,Fabrocinipairing2} the correlated BCS state
\begin{equation}
\ket{\rm CBCS^{alt}}= \sum_{N} F_N\ket{\rm BCS^{(N)}}
\end{equation}
\label{eq:CBCSalt}
where
\begin{equation}
\ket{\rm BCS^{(N)}}=
\sum_m\ket {\Phi_m^{(N)}}
\langle\Phi_m^{(N)} \ket{\rm BCS} \,.
\end{equation}
is the BCS state (\ref{8.6.1}) projected into the $N$-particle
Hilbert space.  This approach leads to a slightly different
gap-equation where the gap function $\Delta_{\bf k}$ under the square
root in Eq. (\ref{eq:gap}) is scaled by a factor
$\bra{\Psi_0}\pver{\kvec}\perz{\kvec}\ket{\Psi_0}^{-2}$, see
Eq. (3.16) od Ref. \onlinecite{HNCBCS}. This factor diverges for
optimized or otherwise long-ranged correlations. To avoid this
divergence we have used here the method developed in
Ref.~\onlinecite{CBFPairing}.

\subsection{Analysis of the gap equation}
\label{ssec:Gapeq}

In the local approximations appropriate for low densities, the
effective interaction is given by Eqs. (\ref{eq:NWloc}).
The pairing matrix element is expressed in terms of the
Fourier-transforms $\tilde {\cal W}(k)$ and $\tilde {\cal N}(k)$:
\begin{equation}
{\cal W}_{\kvec\kvec'}=\frac{1}{N}\tilde{\cal W}(\kvec-\kvec')\,,
\qquad{\cal N}_{\kvec\kvec'}=\frac{1}{N}\tilde{\cal N}(\kvec-\kvec')\,.
\label{eq:Pdef}
\end{equation}
The remaining arguments are standard, {\em cf.\/}
Ref. \onlinecite{FetterWalecka,PethickSmith}: If the gap at the
Fermi surface is small, we can replace the
pairing interaction $\tilde{\cal W}(k)$ by its $s$-wave matrix
element at the Fermi surface,
\begin{equation}
\tilde {\cal W}_F \equiv \frac{1}{2 \KF^2}\int_0^{2\KF} dk k \tilde{\cal W}(k)
= N{\cal W}_{\KF,\KF}\,.
\label{eq:V1S0}
\end{equation}
Then we can write the gap equation as
\begin{align}
1 = - \tilde {\cal W}_F\int\frac{ d^3k'}{(2\pi)^3\rho}
\Bigg[&\frac{1}{\sqrt{(e_{k'}-\mu)^2 + \Delta^2_{\KF}}}\label{eq:gaplowdens}
\\
&- \frac{|e_{k'}-\mu|}{\sqrt{(e_{k'}-\mu)^2 + \Delta^2_{\KF}}}
\frac{\SF(k')}{t(k')}
\Bigg]\nonumber
\end{align}
which is almost identical to Eq. (16.91) in
Ref. \onlinecite{PethickSmith}.  In particular, the second term has
the only function to regularize the integral for large $k'$. We can,
therefore, immediately conclude that the zero temperature gap is, in
this approximation, given by
\begin{equation}
\Delta_{F} = \frac{8}{e^2}\EF\exp\left(\frac{\pi}{ 2a_F\KF}\right)\,.
\label{eq:GapApprox}
\end{equation}
with
\begin{equation}
a_F \equiv   \frac{m}{4\pi\rho\hbar^2}{\cal W}_F\,.
\label{eq:aFdef}
\end{equation}
The low-density limit is then obtained by identifying $a$ with the
vacuum scattering length $\a0$:
\begin{equation}
\Delta_{F}^{(0)} = \frac{8}{e^2}\EF\exp\left(\frac{\pi}{ 2\a0\KF}\right)\,.
\label{eq:GapLowdens}
\end{equation}
Of course, our equations (\ref{5.7.14}), (\ref{eq:gap}) are much more
general: At low densities, the subtraction term -- {\em i.e.\/} the
second term in the square bracket of Eqs. (\ref{eq:Pdef}) and
(\ref{eq:gaplowdens}) is important to regularize the integral for
large momentum transfers. At higher densities, the finite range of the
interaction provides that momentum cutoff and the subtraction term
becomes negligible since the energy numerator term $(|e_k- \mu | +
|e_{k'}- \mu |)$ is zero at the Fermi momentum.

By comparison with the low-density limit and Eq. (\ref{eq:afull}) we
interpret the constant
\begin{equation}
a \equiv \frac{m}{4\pi\rho\hbar^2}\tilde {\cal W}(0+)
\label{eq:amedium}
\end{equation} 
as an ``in-medium'' scattering length.  Hence, at finite densities,
one expects two types of corrections:
\begin{enumerate}
\item[(i)] Medium corrections: The effective pairing interaction
$\tilde {\cal W}(k)$ is related to $\tilde v_{\rm CW}(k)$ through
Eqs. (\ref{eq:FermiRPA0})- (\ref{eq:GFHNC}) which leads to
\begin{eqnarray}
{\cal W}(r) &=&  v_{\rm CW}(r) + (1+\Gamma_{\!\rm dd}(r))w_{\rm I}(r)
\nonumber\\
&=&\left[1+ \Gamma_{\!\rm dd}(r)\right]\left[v(r)+w_{\rm I}(r)\right]\nonumber\\
&&        + \frac{\hbar^2}{m}\Big|\nabla\sqrt{1+\Gamma_{\!\rm dd}(r)}\Big|^2\,.
\label{eq:Wofr}
\end{eqnarray}
Because of Eq. (\ref{eq:vcw0}) and the fact that the induced interaction
$w_{\rm I}(r)$ is of second order in the interaction, we conclude that
that
\begin{equation}
a = \a0\left[1+{\cal O}(\a0\KF)\right]\,.
\end{equation}
In the same order, non-local contributions to the pairing interaction
(\ref{eq:NWloc}) \cite{CBF2} contribute. These can be identified with
particle-hole ladder diagrams and vertex corrections.  Topologically,
one of these diagrams corresponds to the polarization correction
identified by Gorkov {\em et al.\/} \cite{Gorkov}. Moreover, similar
to our analysis of the low-density limit, local correlation functions
will not get the right coefficient of the term proportional to
$(\a0\KF)^2$, hence CBF corrections to the pairing interaction
\cite{CBFPairing} will also lead to modifications of order $(\a0\KF)^2$.

\item[(ii)] The solution of the $s$-wave gap equation is dominated by
  the matrix element (\ref{eq:V1S0}) of ${\cal W}(r)$ at the Fermi
  surface which leads to the solution (\ref{eq:GapApprox}). Only if
  $\tilde{\cal W}(k)$ is practically constant for $0\le k\le 2\KF$, we
  can identify $a_F$ with the in-medium scattering length $a$. The
  dominant finite-range correction to the pairing interaction comes
  from the kinetic energy term
  $\frac{\hbar^2}{m}\left|\nabla\sqrt{1+\Gamma_{\!\rm
      dd}(r)}\right|^2$.  For interparticle distances much larger than
  the interaction range $\sigma$, but smaller than the average
  particle distance, this term is dominated by the vacuum solution of
  the Euler equation, $\sqrt{1+\Gamma_{\!\rm dd}(r)} =
  1-\frac{a_0}{r}$. For interparticle distances larger than $1/\KF$,
  we obtain from Eqs.~(\ref{eq:FermiRPA0}) and (\ref{eq:GFHNC}) that
  $\Gamma_{\!\rm dd}(r)$ falls off like
\begin{equation}
\Gamma_{\!\rm dd}(r) \sim -{9\over 8}{V_{\rm p-h}(0+)\over \hbar^2 \KF^2/2m}
{1\over r^2\KF^2}
\label{eq:GammaLong}\,.
\end{equation}
Consequently, the effective interaction ${\cal W}(k)$ is quadratic in
$k$ for $k\le\KF$ and has a linear dependence \begin{equation}
\tilde{\cal W}(k) = \frac{4\pi\rho a}{m}\left(1-\frac{\pi}{4}a
k\right)\,
\label{eq:Wofk}
\end{equation}
for $k>\KF $. The variation of the pairing interaction between $k=0$
and $k=\KF$ is interaction-dependent and causes, as we shall see in
Sec. \ref{ssec:bcs} a correction that is larger than the ones due to the
effects mentioned above.
\end{enumerate}
All of the corrections discussed above except the one due to the
quadratic momentum dependence of $\tilde{\cal W}(k)$ for $0\le k\le
\KF$ are order $\a0\KF$ and higher {\em i.e.\/} thy lead to a density
dependence of the in-medium scattering length of the form
\begin{equation}
a  = \a0\left[1 + \alpha\frac{\a0\KF}{\pi}+\ldots\right]
\label{eq:aofa0}
\end{equation}
where $\alpha$ is a numerical constant. Among others, the polarization
correction discussed by Gorkov \cite{Gorkov} has this structure.
Inserting the above expansion in (\ref{eq:GapApprox}) changes the
pre-factor to
\begin{align}
\Delta_F &\approx
\frac{8}{e^2}\EF\exp\Bigg(\frac{\pi}{2\a0\KF
\Big(1 + \alpha\displaystyle\frac{\a0\KF}{\pi}\Big)}\Bigg)\nonumber \\
&=
\frac{8}{e^2}\EF\exp\left(-\frac{\alpha}{2}\right)
\exp\left(\frac{\pi}{2\a0\KF}\right)\,.
\label{eq:scaled}
\end{align}
In other words, to the extent that an expansion in powers of
$(\a0\KF)$ is legitimate, all of these corrections just lead to a
modified, but universal, pre-factor in Eq. (\ref{eq:GapLowdens}). This
does not apply to the finite-range correction of the pairing
interaction in the relevant regime $k\le\KF$. One would, of course,
expect that this finite-range correction is of the same order of
magnitude. We shall see in Sec. \ref{ssec:bcs} that its value depends,
on details of the interaction. Hence, we conclude that the exponential
behavior of Eq. (\ref{eq:GapLowdens}) is universal whereas the
pre-factor is not.

\section{Results}
\label{sec:results}

\subsection{Energetics}
\label{ssec:energetics}

We have examined in this paper two model potentials, namely a Lennard-Jones
(LJ) potential
\begin{equation}
V_{\rm LJ} = 4\epsilon \left[\left(\frac{\sigma}{r}\right)^{12}
-\left(\frac{\sigma}{r}\right)^{6}\right]
\label{eq:VLJ}
\end{equation}
and an attractive square well (SW) potential
\begin{equation}
V(r) =
\begin{cases}
-\epsilon & \text{if}\quad r < \sigma\,, \\
\phantom{-} 0&  \text{if}\quad r > \sigma\,.\\
\end{cases}
\end{equation}
Both potentials are parametrized by a characteristic length $\sigma$ and the
depth of the attractive well $\epsilon$.  In both cases, we measure
energies in units of $\hbar^2/2 m \sigma^2$, and length in units of
$\sigma$. Thus, the interaction strength $\epsilon$ and the density
are the only free parameters.

Our choice of interactions provides effective potentials designed to
avoid the instability against clustering that exists for realistic
alkali interactions, but otherwise be close to a realistic
situation. The simplest connection to real interactions is provided by
the vacuum $s$-wave scattering length $\a0$.  The procedure is
legitimate in the low-density limit, many observable properties of
these gases, such as the energy (\ref{eq:lowdensFermi}), depend indeed
only on the $s$-wave scattering length \cite{HuangYang57,Landau5}.
For higher densities this ``universal'' behavior ceases. It is the
purpose of our calculation to explore that area, and also study the
model dependence, by comparing results for the LJ and SW model.  To
make contact with low-density expansions, as well as to determine the
range of ``universal behavior'', we shall use the $s$-wave scattering
length $\a0$ instead of the well-depth $\epsilon$ to characterize the
potential.

\begin{figure}
{\includegraphics[width=1.0\columnwidth]{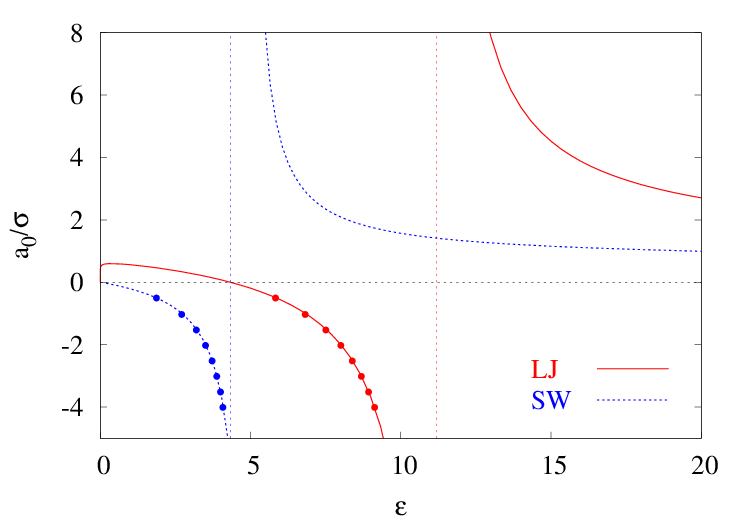}}
\caption{(color online) The plot shows the scattering length $\a0$ as
a function of the interaction strength for the LJ (red, solid) and
the SW (blue, dashed) potential.  The vertical lines (at $\epsilon =
11.18$ for LJ and $\epsilon = 4.336$ for SW) indicate the
interaction strength where a two-body bound state appears. The dots
on the lines indicate the interaction strengths corresponding to
vacuum scattering lengths $\a0/\sigma = -0.5, -1.0,\ldots\-4.0$ for
which we highlight the Landau parameter and the in-medium scattering
lengths in Figs.~\ref{fig:x0hc} and \ref{fig:x0lj}.}
\label{fig:scatplot}
\end{figure}

For the SW potential, $\a0$ is given by
$\a0=\frac{1}{\kappa}(\kappa\sigma - \tan\kappa\sigma)$ where
$\kappa=\sqrt{m\epsilon/\hbar^2}$.  For the LJ potential, $\a0$ must
be obtained numerically.  We show $\a0$ in Fig.~\ref{fig:scatplot} as
a function of the potential well depth $\epsilon$.  The attractive SW
potential has negative scattering length for an interaction strength
below the first resonance. The LJ potential has a positive $\a0$ below
$\epsilon=4.336$, indicating an effectively repulsive interaction.
Thus, when the interaction strength $\epsilon$ of the LJ potential is
raised, starting from $\epsilon=0$, we find three regimes.  (i) For
$0\,<\,\epsilon\,<\,4.336$ we have $\a0\,>\,0$ and there is no bound
state; the many-body ground state is a normal Fermi gas.  (ii) For
$4.336<\epsilon<11.18$ we have $\a0\,<\,0$. There is still no two-body
bound state.  Due to the effectively attractive potential, the
many-body ground state is, at low densities, a BCS state.  (iii) At
$\epsilon\,=\,11.18$ the LJ potential has a resonance at zero
scattering energy.  For $11.18\,<\,\epsilon$, $\a0$ becomes positive
again, and the potential supports at least one two-body bound
state. All these states of Fermi gases have been studied extensively
in experiments with ultracold alkali gases as discussed in the
introduction.  In a previous paper,~\cite{ljium} we have already
studied both, $\a0\,>\,0$ and $\a0\,<\,0$ for fermions, and
$\a0\,>\,0$ for bosons. In that work we have also examined more
sophisticated versions of the FHNC-EL method and have concluded that
these are necessary only at densities comparable to that of liquid
helium. The reader is referred to that work to assess the range of
densities for which the very simple version of the theory spelled out
in Sec. \ref{ssec:FHNC} is reliable. In short, the accuracy of our
energy calculations is expected to be better than 1 percent below a
density of $\rho = 10^{-2}\,\sigma^{-3}$, whereas the error of the
simple FHNC-EL version is about 10 percent as the density increased to
$\rho = 0.4\,\sigma^{-3}$ which is close to the freezing density of
\he3.

We focus in the present work on the effect of attractive interactions
($\a0\,< 0\,$). We have solved
Eqs. (\ref{eq:FermiRPA0})-(\ref{eq:GFHNC}) on a mesh of $2^{18}$
points, with a resolution of 30 points between $r=0$ and $r=\sigma$,
amounting to a box size of 8732$\,\sigma$.  Such a huge box size is
necessary to obtain a reasonable momentum space resolution at the very
low densities we are considering here: Note that a Fermi wave number
of $10^{-3}\sigma^{-1}$ corresponds to a wavelength of $6000\sigma$,
hence this box size is the bare minimum of what one should take to
resolve features of the order of the Fermi wave number. All our
calculations are done for the range of interaction strength where
there is no two-body bound state, {\em i.e.\/} before the first
resonance of $\a0$ appears (indicated by vertical lines in
Fig. \ref{fig:scatplot}).

\begin{figure}
\centerline{\includegraphics[width=0.7\columnwidth,angle=-90]{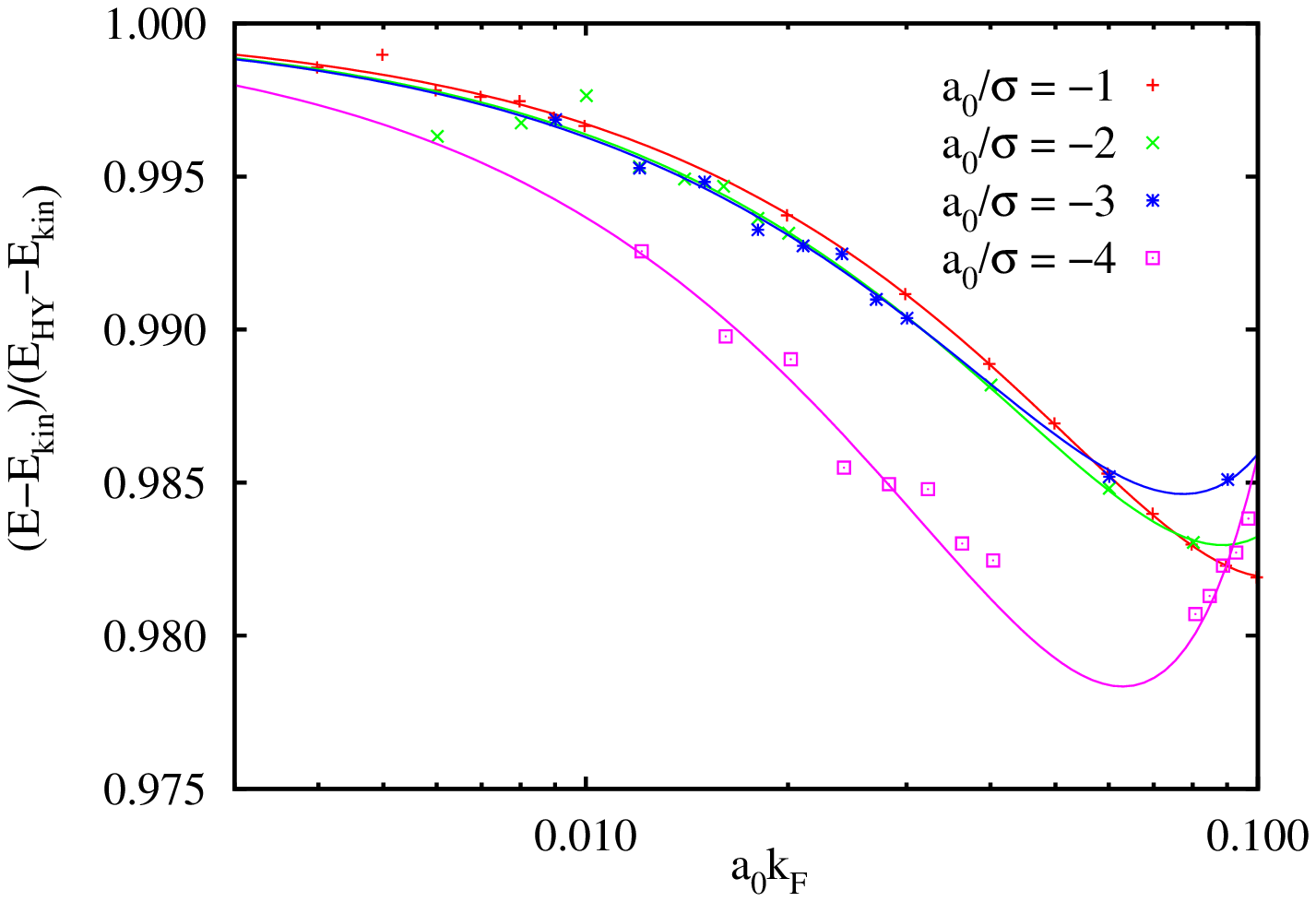}}
\caption{(color online) The plot shows the interaction contribution to
  the equation of state, normalized to the low density expansion, {\em
    i.e.}  the second and third term in the expansion
  (\ref{eq:lowdensFermi}) for the square-well potential.  We show
  results for scattering lengths $ \a0/\sigma = -1,-2,-3,-4$, the
  symbols indicate the numerical values and the curves a second-order
  polynomial fit of the form $E/E_0 = 1+\alpha (\a0\KF) + \beta
  (\a0\KF)^2$. $E_{\rm HY}$ is the expansion (\ref{eq:lowdensFermi}).
\label{fig:eoshc}}
\end{figure}

\begin{figure}
\centerline{\includegraphics[width=0.7\columnwidth,angle=-90]{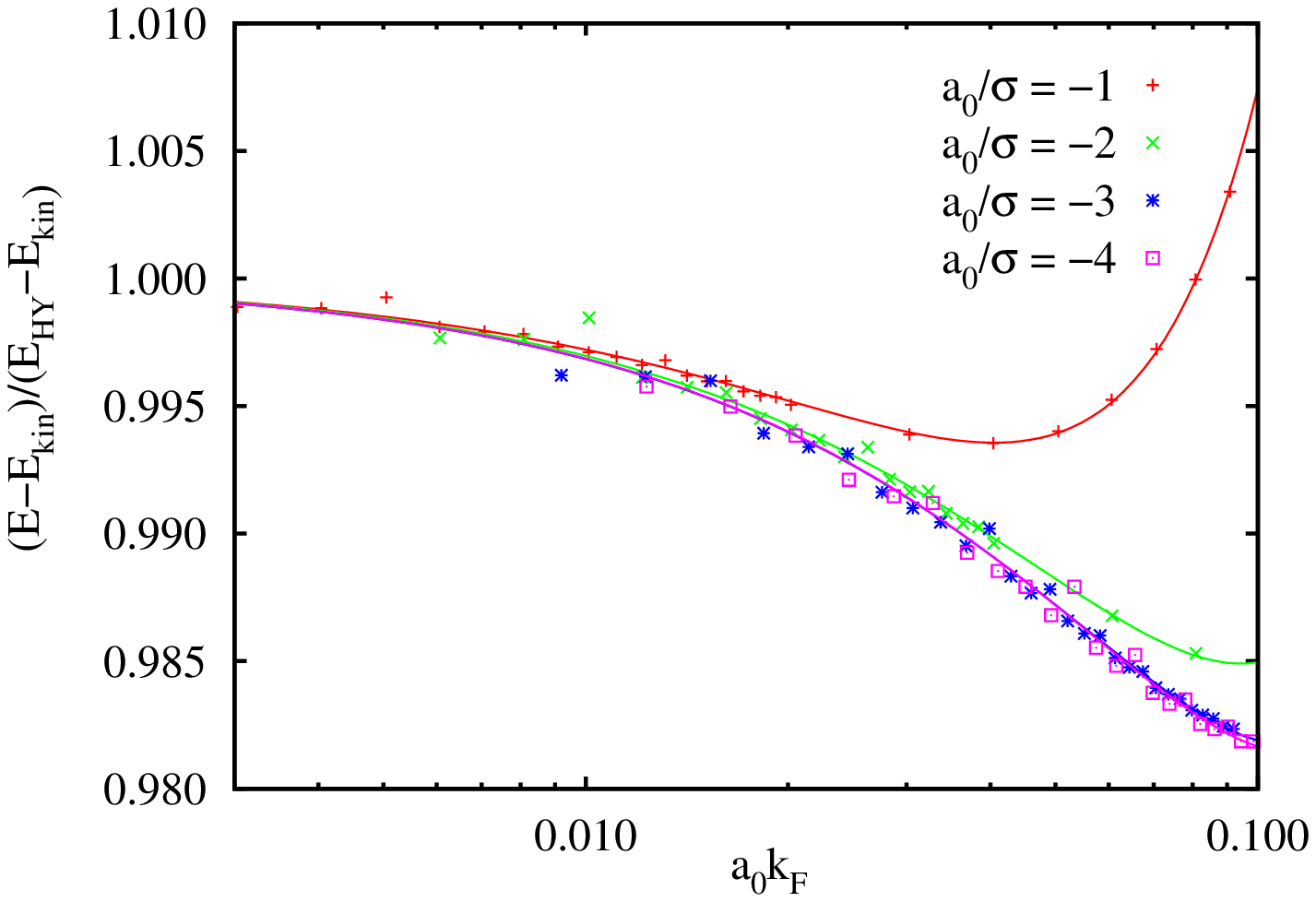}}
\caption{(color online) Same as Fig.~\ref{fig:eoshc} for the Lennard-Jones
model of the interaction.
\label{fig:eoslj}}
\end{figure}

Our equation of state for the two potential models is shown in
Figs. \ref{fig:eoshc} and \ref{fig:eoslj}. To recover the exact
low-density limit (\ref{eq:lowdensFermi}), we have added the
second-order CBF correction (\ref{eq:E2CBF}). To emphasize the
interaction terms, we have subtracted the kinetic energy $E_{\rm kin}
= \frac{3}{5} \EF N$. We have normalized the equation of state to the
expansion (\ref{eq:lowdensFermi}). Thus, Figs.  \ref{fig:eoshc} and
\ref{fig:eoslj} show only the model-dependent correction to the
equation of state.  Omitting the CBF correction (\ref{eq:E2CBF}) and
comparing to the low-density expansion (\ref{eq:lowdensfhnc0}) gives
practically the same results, they are therefore not shown.

Figs. \ref{fig:eoshc} and \ref{fig:eoslj} show already a visible
dependence of the equation of state on the potential model at
approximately $\a0\KF \ge 0.01$ where the third term in the expansion
(\ref{eq:lowdensFermi}) is of the order of $10^{-4}$. In other words,
for both interactions the model-dependent corrections are of the same
order of magnitude as the third term in the expansion
(\ref{eq:lowdensFermi}). For both interactions we observe that,
dependent on the interaction strength, the equation of state begins to
deviate strongly from a simple, smooth power law for $\a0\KF > 0.02$.

The most interesting feature we observe is that the FHNC--EL equations
cease to have solutions at sufficiently large values of the density or
of $-\a0$.  Such a limit is expected for sufficiently attractive
interactions: The Fermi gas is, in the low density limit, stabilized
by the Pauli pressure. As the density increases, the energy per
particle becomes negative and the static incompressibility
\begin{equation}
mc^2 = \frac{d}{d\rho}\rho^2\frac{d}{d\rho}\frac{E}{N}\,,
\label{eq:mc2}
\end{equation}
where $c$ is the hydrodynamic speed of sound, goes to zero.
Such an effect has already been reported by Owen \cite{OwenVar}.
$mc^2\to 0$ indicates a spinodal instability, where the system
separates into a low and a high density phase.  On the other
hand, it is widely accepted that a low density two-component Fermi gas
is subject to dimerization close to the unitary limit, $\a0\to -\infty$.
In the following we will argue that dimerization indeed occurs,
but well before the limit $\a0\to -\infty$.  Instead,
dimerization is accompanied by the divergence of the in-medium scattering
length.

Let us examine the question of stability from the point of view of the
existence of solutions of the FHNC-EL equations: {\em In general,\/}
the FHNC-EL equations cease to have solutions if the assumed wave
function is unstable against small perturbations. This is most clearly
seen for the case of density fluctuations: The term under the square
root of Eq. (\ref{eq:FermiRPA0}) must be positive. In the limit
$k\rightarrow 0+$ this leads to the condition
\begin{equation}
1+F_0^s \equiv
1 + \frac{4 m}{\hbar^2 \KF ^2}\left(\frac{3}{4}\right)^2 \tilde V_{\rm p-h}(0+)
\rightarrow 1 + \frac{3}{\pi}\a0 \KF  > 0.
\label{eq:mc2coll}
\end{equation}
The right-most expression is the low density limit, $\a0\KF\ll 1$, where
$F_0^s$ is small and the in-medium scattering length $a$ is well
approximated by the vacuum scattering length $\a0$, see Eq. (\ref{eq:mc2full}).
The limit can be regarded as the low density limit of the particle-hole
interaction,
\begin{eqnarray}
\tilde V_{\rm p-h}(0+)
&=& \tilde v_{\rm CW}(0+) + \rho\int d^3r \Gamma_{\!\rm dd}(r)w_{\rm I}(r)
\label{eq:Vphrho}\\
& \rightarrow& \frac{4\pi\rho\hbar^2}{m}\,\a0
\quad\mbox{as}\quad\rho\rightarrow 0\,.  \nonumber
\end{eqnarray}
We have used above the fact that the induced interaction is of second
order in the bare interaction.  Hence, the system can be driven into
an instability by holding the potential fixed and simply increasing
the density factor in Eq. (\ref{eq:Vphrho}). Note that the above
stability limit (\ref{eq:mc2coll}) is valid for the local correlation
operator (\ref{eq:wavefunction}). In an improved calculation that does
not rely on a local correlation operators but rather includes CBF
corrections to all orders \cite{polish} the stability condition would
read
\begin{equation}
1 + \frac{3 m}{\hbar^2 \KF ^2} \tilde V_{\rm p-h}(0+)
= 1 + \frac{2}{\pi}a \KF  > 0\,.
\label{eq:mc2full}
\end{equation}
The condition is immediately recognized as the stability condition of
Landau's Fermi Liquid theory, $F_0^s > -1$. Note that the ground state
theory formulated in Eqs. (\ref{eq:FermiRPA0})-(\ref{eq:GFHNC}) does
not contain self-energy insertions (called ``cyclic-chain'' diagrams
in the FHNC theory), which means that the effective mass is equal to
1. According to our findings in Ref. \onlinecite{ljium}, this is an
acceptable approximation at the low densities under consideration
here. Since the interaction contribution to $F_0^s$ is, at low
densities, proportional to $k_F$, an instability against density
fluctuations could occur for sufficiently attractive
interactions. This has the consequence that the FHNC-EL equations
cease to have a solution.

\begin{figure}
\centerline{\includegraphics[width=0.7\columnwidth,angle=-90]{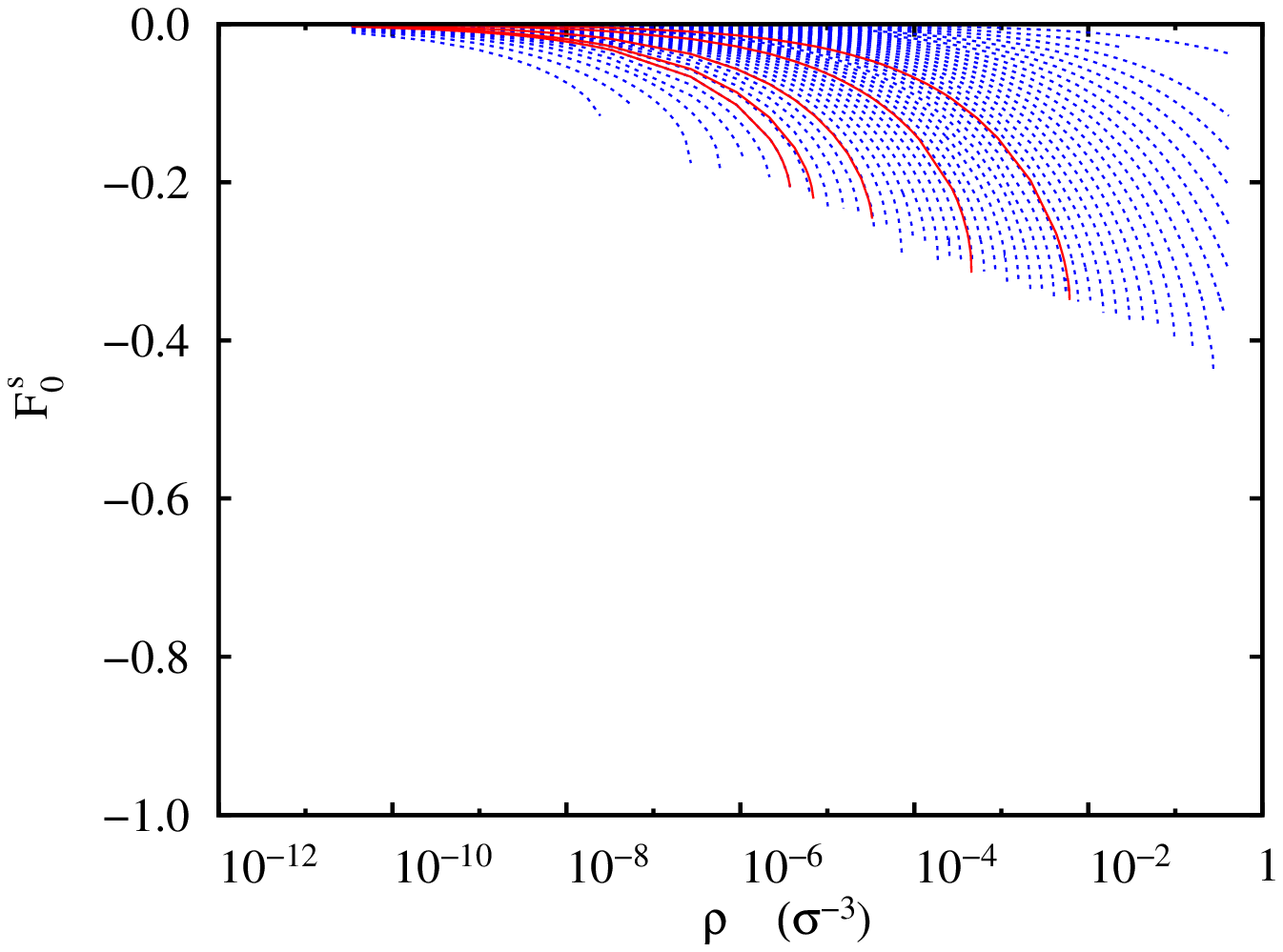}}
\caption{(color online) The plot shows the dependence of the Landau
parameter $F_0^s$ for the attractive square-well potential as a
function of the density for a sequence of coupling strengths
$\epsilon = 0.1, 0.2, \ldots, 4.6$ (blue, dashed curves) and vacuum
scattering lengths $\a0/\sigma = -0.5,-1.0,\ldots,-4.0$ (red, solid
curves) that correspond to the dots in Fig. \ref{fig:scatplot}.  The
curve that ends at the lowest density corresponds to the strongest
interaction $\epsilon = 4.6,\ \a0/\sigma=-11.2\,$, whereas the ones
corresponding to the weak interactions $0.1\le\epsilon\le 0.7$ are
stable beyond a density of $\rho=0.4\sigma^{-3}$.
\label{fig:x0hc}}
\end{figure}

\begin{figure}
\centerline{\includegraphics[width=0.7\columnwidth,angle=-90]{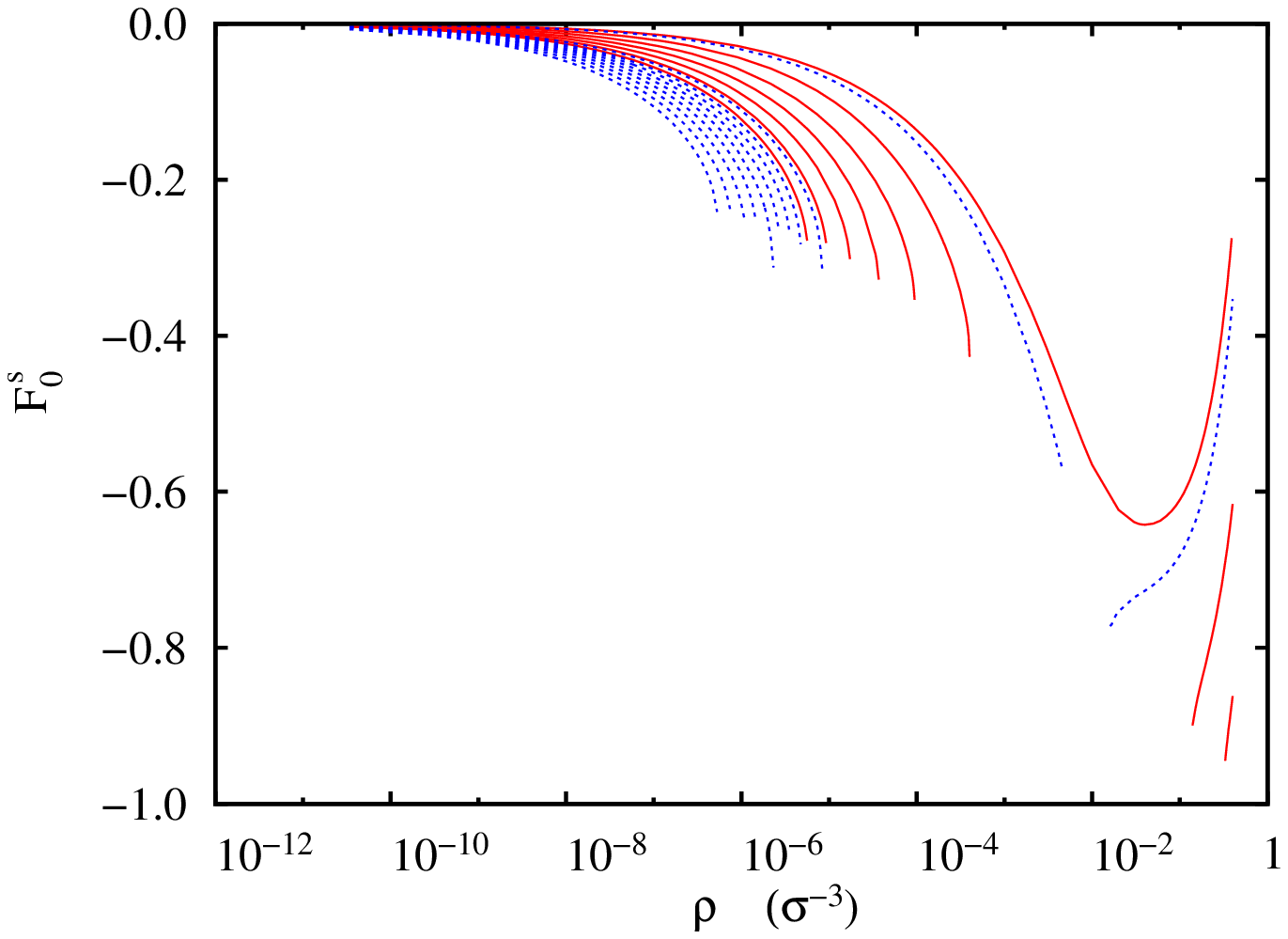}}
\caption{(color online) Same as Fig. \ref{fig:x0hc} for the
  Lennard-Jones model of the interaction. The blue (dashed) lines
  correspond to coupling strengths $\epsilon = 7.0, 9.0, 9.2,
  9.3,\ldots 9.9$.  Note that the Lennard-Jones model also supports a
  high-density condensed phase: The curves for the interaction
  strengths $\epsilon = 7.0\ (\a0/\sigma=-1.12)$ (blue, dashed),
  $\epsilon = 7.51\ (\a0/\sigma=-1.5)$ and $\epsilon =
  8.01\ (\a0/\sigma=-2)$ (both red, solid) are discontinuous at high
  density. The curve that ends at the lowest density corresponds to
  the strongest interaction.
\label{fig:x0lj}}
\end{figure}

We found indeed an instability of the solutions of the FHNC-EL
equations. However, this instability does {\em not\/} appear to be
caused by $F_0^s \rightarrow -1$. In fact, in our calculations we were
not able to come close to the point of spinodal instability. We show
in Figs.~\ref{fig:x0hc} and \ref{fig:x0lj} the value of $F_0^s$ as a
function of density for a family of potential strengths.  For the SW
potential, we have, depending on the coupling strength $\epsilon$,
been able to solve the FHNC-EL equations in the regime between
$\rho=3.4\times 10^{-11}\sigma^{-3}$ up to a critical density that
depends on the coupling strength. At that density, $F_0^s$ begins to
drop very rapidly but it does not appear to approach the critical
value of $-1$ for a spinodal instability, see Fig.~\ref{fig:x0hc}.  We
also note that an instability signified $F_0^s\rightarrow -1$ means a
transition to a state with non-uniform density which is not what has
been observed experimentally.

Whereas a system of fermions interacting with an attractive SW
potential exhibits an instability as the density is increased, the LJ
model interaction leads, due to its repulsive core, to a richer phase
diagram that also features a high-density condensed phase -- the
liquid phase of \he3. At low densities, we see the same picture as for
the SW potential: The liquid is stabilized by the Pauli pressure, but
the FHNC-EL equations cease to have solutions above a certain density
where the Landau-parameter $F_0^s$ is still far from its critical
value $F_0^s=-1$. The situation is different in the high-density
regime: For sufficiently strong interactions, the system also can
develop a high-density condensed phase. The noteworthy feature is that
one can get much closer to the spinodal instability limit $F_0^s=-1$
from higher densities than from lower densities.

\begin{figure}
\centerline{\includegraphics[width=1.0\columnwidth]{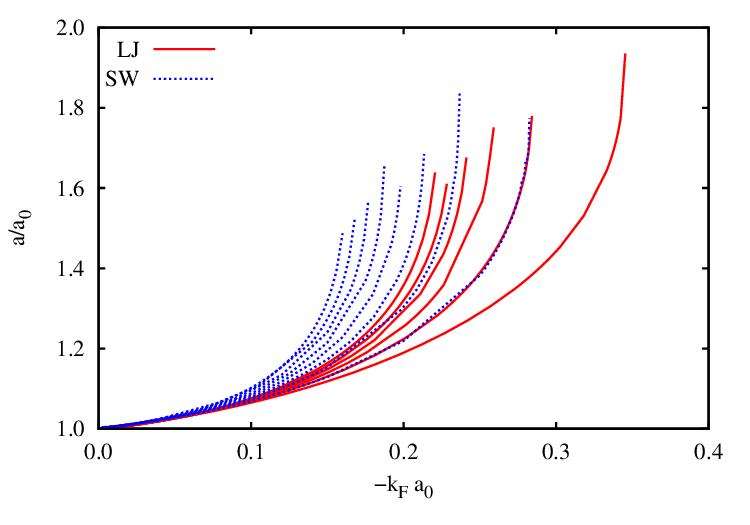}}
\caption{(color online) The ratio between the in-medium scattering
length $a$, Eq. (\ref{eq:amedium}), and the vacuum scattering length
$\a0$ as function of $-\KF\a0$, for both the LJ (full line) and the
SW potential (dashed line).  The different curves correspond to
different values of $\a0/\sigma=-0.5,-1.0,\dots,-4.0$ (SW) and
$\a0/\sigma=-1.5,-2.0,\dots,-4.0$ (LJ), with the higher curves
corresponding to larger $|\a0|$.  The deviation of $a/\a0$ from unity
is universal only for low $\KF$.  The divergence of $a/\a0$
indicates dimerization.
\label{fig:amed_all}}
\end{figure}

\begin{figure}
\centerline{\includegraphics[width=1.0\columnwidth]{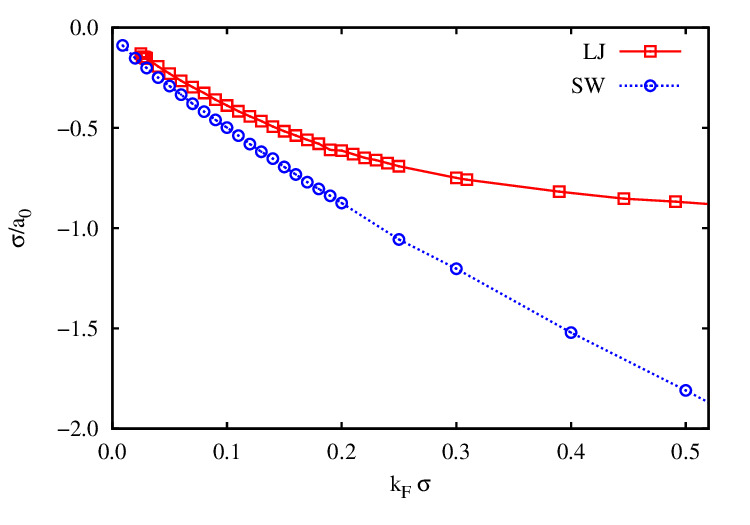}}
\caption{(color online) The figure shows, for the square-well (blue
  line with circle markers) and the Lennard-Jones (red line with
  circular markers) potential models, the inverse of the critical
  value of the vacuum scattering length $\a0$ below which a
  non-dimerized Fermi liquid phase exists.
\label{fig:acrit}}
\end{figure}

To examine the nature of the instability, we show in
Fig.~\ref{fig:amed_all} the in-medium scattering length $a$, see
Eq.~(\ref{eq:amedium}).  The ratio $a/\a0$ is shown as function of
$-\KF\a0$ for increasing values of $|\a0|$: $\a0/\sigma=-0.5,-1.0,\dots,-4.0$
for the SW model and $\a0/\sigma=-1.5,-2.0,\dots,-4.0$ for the LJ
model.  We do not show $\a0/\sigma=-0.5$ and $-1.0$ for the LJ model,
because the system is stable for all densities as discussed
above. Similar to the vacuum scattering length, the in-medium
scattering length exhibits a singularity. The location of the
singularity depends obviously on both the density and the interaction
model. Evidently, medium corrections to the effective interactions
$\tilde{\cal W}(k)$ have the effect that the in-medium scattering
length exhibits a singularity as $\KF$ is increasing.  The larger
$|\a0|$, the smaller the critical $\KF$ value where the the divergence
happens, see also Fig. \ref{fig:acrit}.  $a/\a0$ is universal only for
very low densities, where all curves merge into a single curve
converging towards unity in the zero density limit.  For finite
density, $a/\a0$, and thus $\KF a$, depends on $\a0$, $\KF$, and the
interaction model, and not just on the dimensionless parameter
$\KF\a0$.

The appearance of such an effect is not surprising because the leading
correction to the bare interaction is the attractive phonon-exchange.
This leads, as the density is increased, to a divergence of the
in-medium scattering length $a$. Thus, we conclude that the
instability found in our calculations is an indication of
phonon-induced dimerization. Further evidence for the appearance of a
phonon-exchange induced dimerized phase will be provided in the next
section where we discuss distribution functions. 

Figs.~\ref{fig:amed_all} show essentially the same scenario for the SW
and the LJ model. At a given density, the critical scattering length
$\a0$ where dimerization occurs is model-dependent.  But the
difference becomes smaller as the density is reduced.  In order to
determine the critical scattering length, we have extrapolated the
in-medium scattering length to the point where it diverges. The result
for both interactions are shown in Fig. \ref{fig:acrit}. There we show
the inverse of the critical scattering length where the normal state
becomes unstable as function of $\KF\sigma$.  The curves for the LJ
and the SW potential approach each other when $\KF\sigma$ is well
below 0.1.  This shows that the critical value of $\sigma/\a0$ becomes
universal in the low density limit. As a consequence of many-body
effects, this critical value of $\sigma/\a0$ is not zero.

The physical message that emerges here is actually quite simple:
Normally the transition from a ``BCS'' state to a ``BEC'' state is
discussed in terms of the vacuum scattering length $a_0$; it is clear
that, as $a_0$ diverges, the expressions (\ref{eq:GapLowdens}) or
(\ref{eq:scaled}) for the gap become meaningless. To be rigorous, one
should however use in these equations the in-medium scattering length
$a$ and not the vacuum scattering length $a_0$. We have justified this
(see Eq. (\ref{eq:Wofr})), we have calculated the in-medium scattering
length, and found a divergence of $a$ {\em before\/} the vacuum
scattering length $a_0$ diverges. This is expected because many-body
effects change the interaction, and the first correction term is
attractive.

Similar dimerization effects have been observed in two-dimensional
$^3$He-$^4$He mixtures, in double-layers of bosonic dipoles
\cite{PhysRevA.87.033624}, and also in the process of
$\alpha$-clustering of nuclear matter \cite{alpha,alpha2}. We hope
that our result that there are clear many-body effects to be
identified in the BCS regime will encourage experimental
investigations in this regime.

\subsection{Distribution Functions}
\label{ssec:dists}

\begin{figure}
\centerline{\includegraphics[width=1.0\columnwidth]{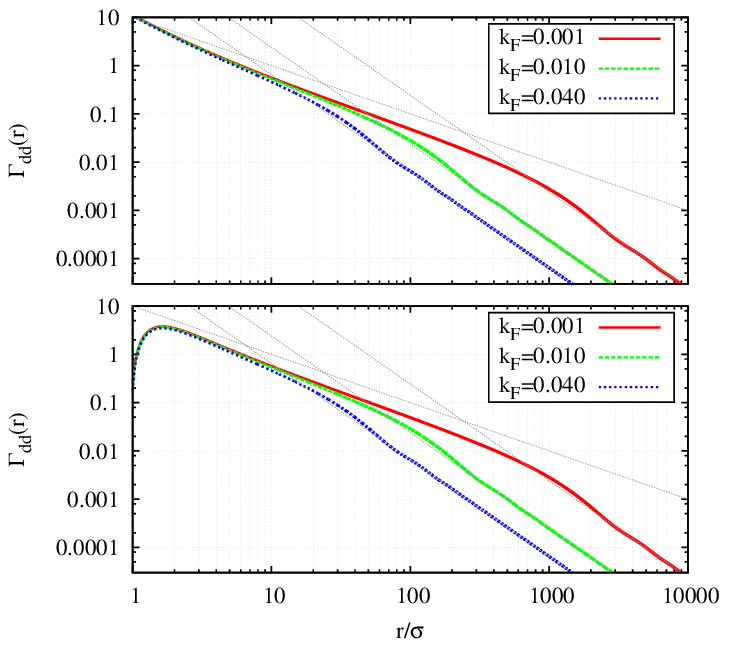}}
\caption{(color online) The direct-direct correlation function
$\Gamma_{dd}(r)$ for $\a0/\sigma=-2.5$ and $\KF\sigma=0.001, 0.01,$
and $0.04$ for the LJ model (upper pane) and the SW model (lower
pane).  Straight lines indicate the asymptotic $1/r^2$ behavior
given by Eq. (\ref{eq:GammaLong}) and another straight line
indicates the $1/r$ behavior for intermediate distances.
\label{fig:glj839hc372}}
\end{figure}

The pair distribution functions for parallel and anti-parallel spins,
$g_\qupup(r)$ (\ref{eq:guu}) and $g_\qupdown(r)$ (\ref{eq:gud}),
contain information about correlations due to Fermi-statistics and,
more interestingly, due to interactions.  The latter are captured by
the direct-direct correlation function $\Gamma_{dd}(r)$, which we show
in Figs.~\ref{fig:glj839hc372} for the LJ and SW model, respectively.
The respective potential strengths were chosen such that the
scattering length was $\a0=-2.5\,\sigma$.  $\Gamma_{dd}(r)$ is shown
for three Fermi wave numbers $\KF\sigma=0.001, 0.01,$ and $0.04$.  One
can discern two regimes: the asymptotic regime where $\Gamma_{dd}(r)$
falls off as $1/r^2$ for $rk_F\gtrsim 1$ due to many-body effects, see
Eq.~(\ref{eq:GammaLong}); and an intermediate regime $\KF\sigma
\lesssim r\KF\lesssim 1$ where $r$ is smaller than the average
particle distance and $\Gamma_{dd}(r)$ falls off like $1/r$ as
expected from two-body scattering in vacuum.  The behavior in the
asymptotic regime can be obtained from the $k\to 0$ limit of
Eqs. (\ref{eq:FermiRPA0}) and (\ref{eq:GFHNC}) which leads to
Eq. (\ref{eq:GammaLong}). In the low-density limit, the speed of sound
is obtained from the equation of state (\ref{eq:lowdensFermi}). Since
only the speed of sound enters, the asymptotic form of
$\Gamma_{dd}(r)$ is independent of the interaction model.  In
Figs.~\ref{fig:glj839hc372} this asymptotic behavior is illustrated by
straight lines.

For low densities, the spin-parallel pair distribution function
$g_\qupup(r)$ is dominated by Fermi statistics.  The Pauli exclusion
principle ensures that $g_\qupup(0)=0$, regardless of interactions.
This is guaranteed by the statistical factor $1-\ell^2(r\KF )$ in
Eq. (\ref{eq:guu}) which suppresses $g_\qupup(r)$ for $r\KF\lesssim
1$, thus screening the interaction.  Note also that
$\lim_{r\rightarrow 0}\left[C(r) +(\Delta \tilde X_{\rm
    ee})_1(r)\right] = 0$. Therefore, the interaction effectively
plays no role in a dilute gas of spin-polarized fermions, which has
also been established in many experiments.  For example for $\KF\sigma
=0.01$, the curves for $g_\qupup(r)$ obtained for the two interactions
are almost indistinguishable from each other and from the distribution
function of non-interacting fermions for a wide range of values of
the vacuum scattering length between $\a0=-0.5$ and $-4.0$.  $g_\qupup(r)$ is
essentially identical to the spin-parallel pair distribution of free
fermions $g^F_\qupup(r)$, which has also been verified by QMC
results~\cite{astraPRL04,morrisPRA10}. Thus, little or no non-trivial
information can be obtained from $g_\qupup(r)$, and we therefore
refrain from further discussing or showing this quantity.

\begin{figure}
{\includegraphics[width=1.0\columnwidth]{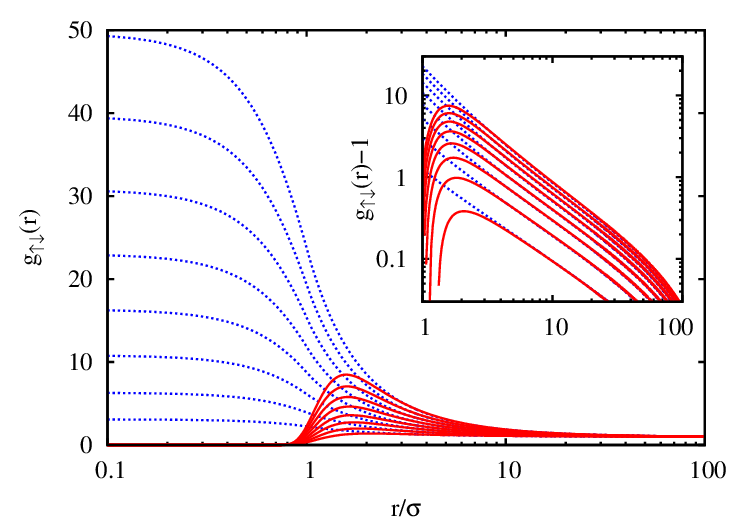}}
\caption{(color online) $g_\qupdown(r)$ for $\KF =0.01/\sigma$ and
  $\a0/\sigma=-0.5, -1.0, -1.5, \ldots -4.0$\,.  The uppermost curve
  corresponds to the largest value of $|\a0|$.  We show results for
  the LJ (solid red lines) and SW (dashed blue lines) potential.  The
  inset shows $g_\qupdown(r)-1$ on a logarithmic scale to illustrate
  that for $r\gtrsim 3\sigma$, $g_\qupdown(r)-1$ is independent of the
  interaction model.}
\label{fig:gant}
\end{figure}

The anti-parallel pair distribution function $g_\qupdown(r)$ is not
suppressed by statistical correlations. Instead, $g_\qupdown(r)$ is
dominated by correlation effects, which are described by the
direct--direct correlation function $\Gamma_{\!\rm dd}(r)$,
Eq.~(\ref{eq:GFHNC}).  The effectively attractive interaction,
$\a0\,<\,0$, leads to an enhancement of $g_\qupdown(r)$ as $r$ is
reduced.  We show in Fig.~\ref{fig:gant} $g_\qupdown(r)$ for $\KF\sigma
=0.01$ and $\a0/\sigma=-0.5, -1.0,\ldots -4.0$, for both the SW and
the LJ potential.  For small distances $r$, $g_\qupdown(r)$ is
dominated by the interaction potential, leading to a large value at
$r=0$ for the SW potential: The maximum value of $g_\qupdown$ is
attained at $r=0$. For the largest $|\a0|$ considered here, we have
obtained solutions of the FHNC-EL equations where $g_\qupdown(0)$ is
more than 150 times larger than $g_\qupdown(\infty)$. The short range
repulsion of the LJ interaction suppresses the pair distribution for
$r<\sigma$, and therefore $g_\qupdown(r)\rightarrow 0$ as $r
\rightarrow 0$. The maximum value of $g_\qupdown(r)$, which is
attained at finite $r$, is therefore much lower than in the SW case.

One might argue that both models are unrealistic at short $r$, the
shape of $g_\qupdown(r)$ for $r\lesssim\sigma$ is of only theoretical
interest. However, we will show in Section \ref{ssec:bcs} that
finite-range effects have a quite visible effect on the superfluid
energy gap and, hence, on the condensation energy.  We also observe
that for $r\gtrsim 3\sigma$, $g_\qupdown(r)$ becomes universal in the
sense that it is independent of the choice of interaction.  This is
not a feature special to cold gases: The asymptotic behavior of
$g_\qupdown(r)$ is determined by the speed of sound
\cite{FeenbergBook}, see also Eq. (\ref{eq:GammaLong}).
$g_\qupdown(r)$ is still strongly enhanced compared to the asymptotic
limit $g_\qupdown(r\to\infty)=1$.  Furthermore we find that
$g_\qupdown(r)$ is very similar to the zero-energy scattering solution
of the two-body Schr\"odinger equation, this is simply due to the fact
that $g_\qupdown(r)$ is dominated by the dynamic correlation function
$1+\Gamma_{\!\rm dd}(r)$, {\em cf.} Eq. (\ref{eq:gud}).  The situation
changes again as we approach the dimerization instability: The maximum
value of $1+\Gamma_{\!\rm dd}(r)$ starts to rise rapidly indicating an
approaching singularity. The effect becomes stronger and appears at
lower density as the coupling strength of the underlying bare
interaction is increased.

\begin{figure} {\includegraphics[width=1.0\columnwidth]{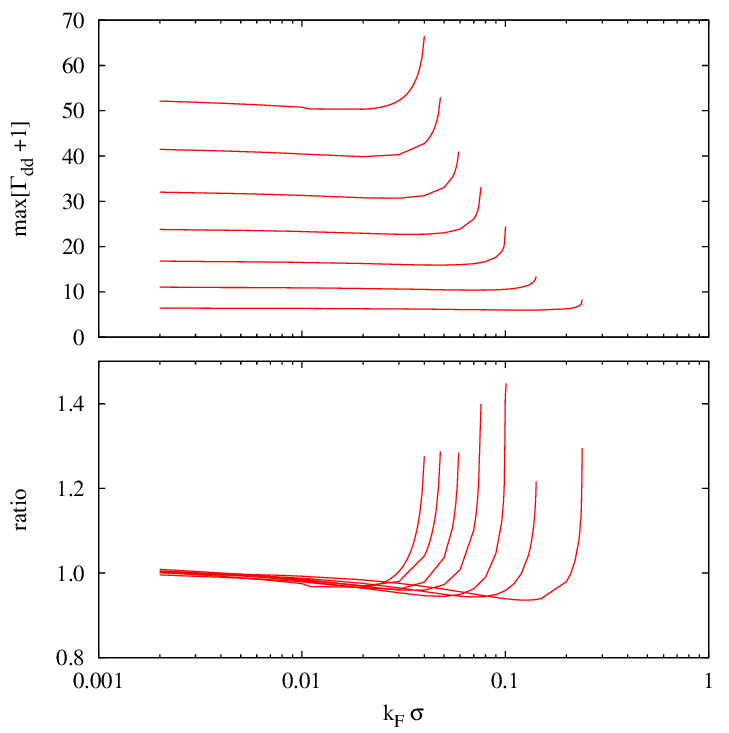}}
\caption{(color online) The upper pane shows, for the square-well
potential, the maximum value of $1+\Gamma_{\!\rm dd}(r)$ for
vacuum scattering lengths $\a0/\sigma = -1.0, -1.5, \ldots
-4.0$. The uppermost curve, which ends at the smallest value of
$\KF\sigma$, corresponds to the largest value of $\a0$. The lower
pane shows the ratio between the the maximum value of
$1+\Gamma_{\!\rm dd}(r)$ and the maximum value of the the vacuum
solution $|\psi(r)|^2$, {\em cf.} Eq. (\ref{eq:scatteq}).}
\label{fig:ghcmax}
\end{figure}

\subsection{BCS pairing}
\label{ssec:bcs}

We now return to the question of the density-- and momentum dependence
of the pairing interaction $\tilde{\cal W}(k)$. We have discussed
already above that the in-medium scattering length can become singular
due to phonon-exchange correction, the considerations of this section
apply therefore only to the case of weak pairing, in other words to a
physical situation where the wave function (\ref{eq:wavefunction}) is
a good approximation.

\begin{figure}
{\includegraphics[width=1.0\columnwidth]{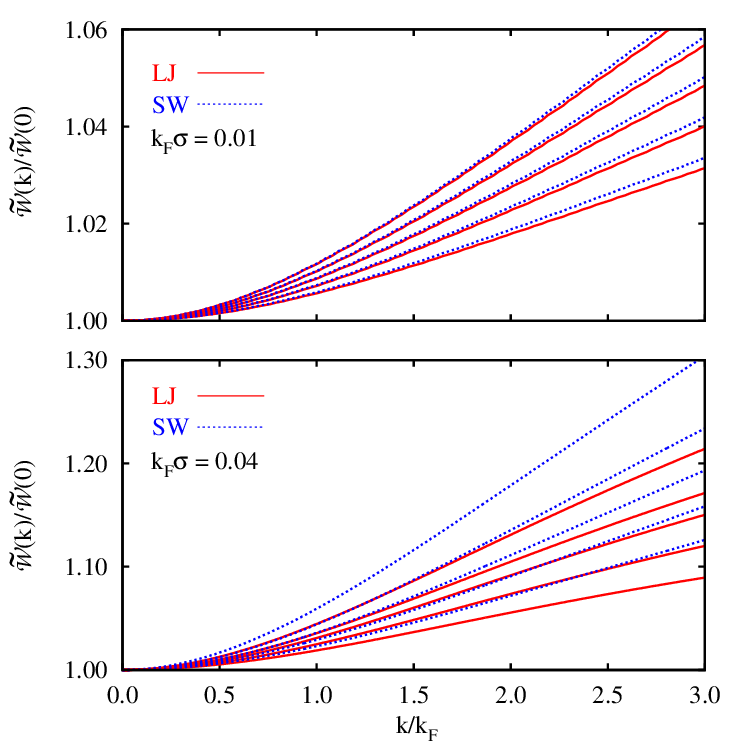}}
\caption{(color online) The figures show the pairing interaction
$\tilde {\cal W}(k)$, normalized to its value at $k\to 0$, for a
sequence of vacuum scattering lengths $\a0/\sigma = -2.0, -2.5,
-3.0, -3.5, -4.0$ and both potential models (full line: LJ, dashed
line: SW) for $\KF\sigma=0.01$ (top panel) and
$\KF\sigma=0.04$ (bottom panel); larger values of $|\a0|$
correspond to higher curves. The linear regime for
$\KF=0.01\sigma^{-1}$ agrees well with the slope predicted by
Eq. (\ref{eq:Wofk}). }
\label{fig:vpairofk}
\end{figure}

Before we describe our calculations, we go back to the momentum
dependence of the pairing interaction.  We show in
Figs. \ref{fig:vpairofk} $\tilde{\cal W}(k)/\tilde{\cal W}(0+)$ for
two values of $\KF$, $\KF\sigma=0.01$ (top panel) and $\KF\sigma=0.04$
(bottom panel). For small $\KF$ such as $\KF\sigma=0.01$ and less, the
regime where $\Gamma_{\!\rm dd}(r)$ behaves as $\sqrt{1+\Gamma_{\!\rm
    dd}(r)} = 1-\frac{a_0}{r}$ is rather large. Therefore, the linear
regime in $\tilde{\cal W}(k)$ is well defined and agrees well with the
prediction (\ref{eq:Wofk}).  For $\KF\sigma=0.04$, the regime where
the $a_0/r$ behavior of the correlations is visible is much smaller,
hence the linear regime is less well defined and $\tilde{\cal W}(k)$
also becomes model dependent. The important point to be made here is,
however, that in neither case, and especially as $|\a0|$ gets larger, is
the simple estimate $\tilde{\cal W}(k)\approx\tilde{\cal W}(0+)$ from
low-density expansions valid for a significant portion of the
integration regime $0\le k\le 2\KF$ which is needed for the
calculation of the $s$-wave pairing matrix elements, see
Eq.~(\ref{eq:V1S0}).

\begin{figure}
{\includegraphics[width=1.0\columnwidth]{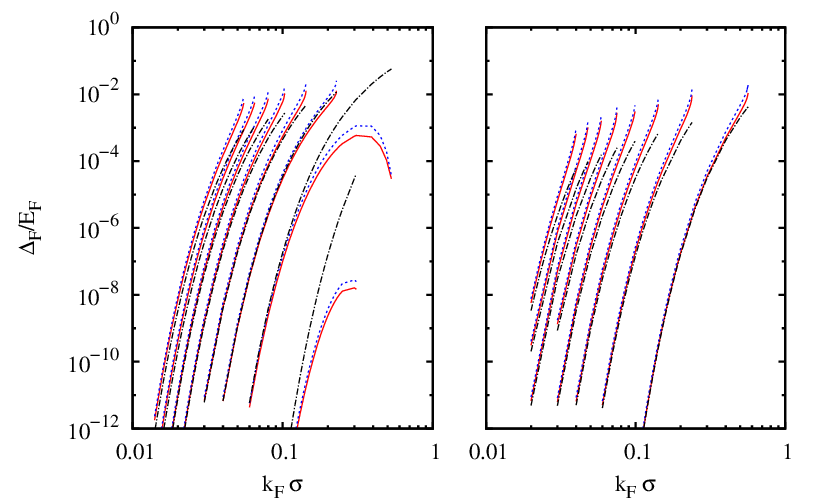}
}
\caption{(color online) The gap $\Delta_{\KF}$ at the Fermi momentum
  in units of the Fermi energy as a function of Fermi wave number
  $\KF$ for the SW model (left panel) and the LJ model (right panel),
  for $\a0/\sigma = -0.5, -1.0, \ldots -4.0$ (higher values of
  $\Delta_{\KF}$ correspond to larger values of $|\a0|$).  Full line:
  full numerical solution; dashed line: approximate solution
  $\Delta_F$ (\ref{eq:GapApprox}); dash-dotted line: Solution
  $\Delta_F^{(0)}$ obtained by setting $a=\a0$ in
  Eq. (\ref{eq:GapLowdens}).  }
\label{fig:GapSWLJbare}
\end{figure}

We have calculated the gap in the excitation spectrum at the Fermi
surface, $\Delta_{\KF}$ (see Eq.~(\ref{eq:gap})), for the LJ and the SW
interaction model, for a wide range of densities, characterized by the
Fermi wave number $\KF $, and $s$-wave scattering lengths $\a0$. A
first overview is shown in Fig.~\ref{fig:GapSWLJbare}, where the higher
values of $\Delta_F$ correspond to larger values of $|\a0|$.

\begin{figure}
{\includegraphics[width=0.7\columnwidth,angle=-90]{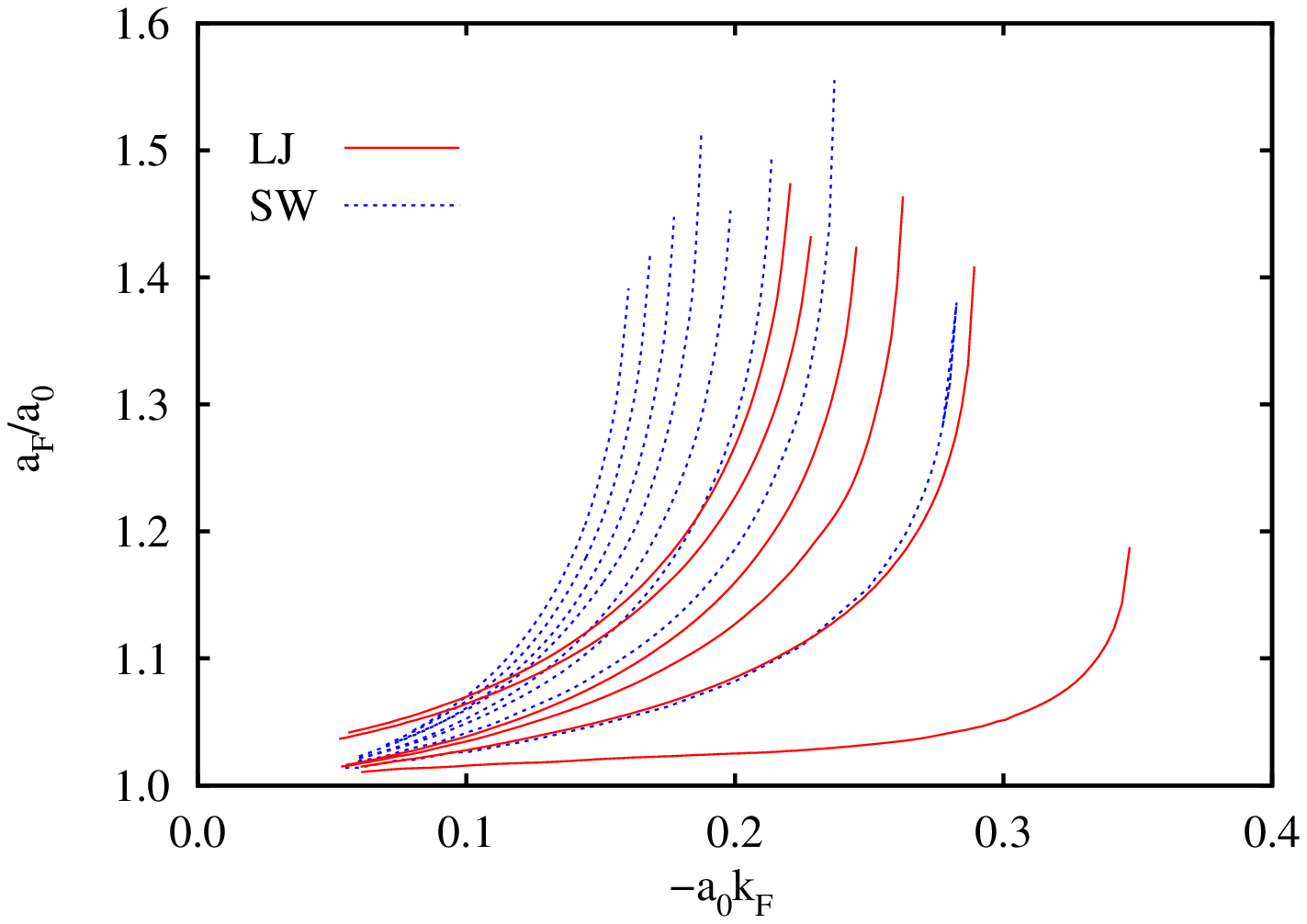}}
\caption{The figure shows the ratio of the scaled pairing matrix
  element $a_F$, Eq. (\ref{eq:aFdef}), to the vacuum scattering length
  $\a0$ for the Lennard-Jones model (red, solid curves) and the
  soft-core interaction model (blue, dashed curves) as a function of
  $-\a0\KF$.  The different curves are for different values of
  $\a0/\sigma=-0.5,-1.0,\dots,-4.0$ (SW) and
  $\a0/\sigma=-1.5,-2.0,\dots,-4.0$ (LJ), with the higher curves
  corresponding to larger $|\a0|$. }
\label{fig:VSCkfkf}
\end{figure}

Figs. \ref{fig:GapSWLJbare} provide two pieces of information: One is
an assessment of the accuracy of the approximate solution
(\ref{eq:GapApprox}). This is obviously quite good throughout the
whole regime of interaction strengths and densities. The major
difference comes, at high densities, from the deviation of the vacuum
scattering length $a_0$ from $a_F$ defined in
Eq. (\ref{eq:aFdef}). The general dependence of the gap on both the
interaction strength and the density is quite similar, in particular
the exponential dependence on the scattering length holds over many
orders of magnitude. This is consistent with the general feature
spelled out in Sec. \ref{ssec:Gapeq} that medium-- and finite--range
corrections are manifested in the pre-factor in
Eq. (\ref{eq:scaled}). This pre-factor is universal as long as the
in-medium scattering length is of the form (\ref{eq:aofa0}) which is
the case, among others, if the momentum dependence of $\tilde{\cal
  W}(k)$ is linear.  Deviations from this universal behavior can only
be expected from the quadratic behavior of $\tilde{\cal W}(k)$ for
$0\le k \lesssim\KF$ which is a direct manifestation of many-body
effects. Indeed, this correlation effect is significant: We show in
Fig. \ref{fig:VSCkfkf} the ratio $a_F/\a0$, see Eq.~(\ref{eq:aFdef}),
in the density/interaction strength regimes where the gap is larger
than $10^{-10}\EF$. Evidently, the behavior is, in that regime,
neither linear, nor a universal function of $a_0\KF$. Only at very
small values of $\a0\KF$ where the gap is of the order of
$10^{-8}\EF$, a linear behavior might be interpreted, but the slope of
such a linear behavior depends on the interaction model. Calculations
at even lower density where a linear regime might be found require a
much bigger mesh than the one used here to reliably resolve the region
$0\le k \lesssim\KF$.

\begin{figure}
{\includegraphics[width=1.0\columnwidth]{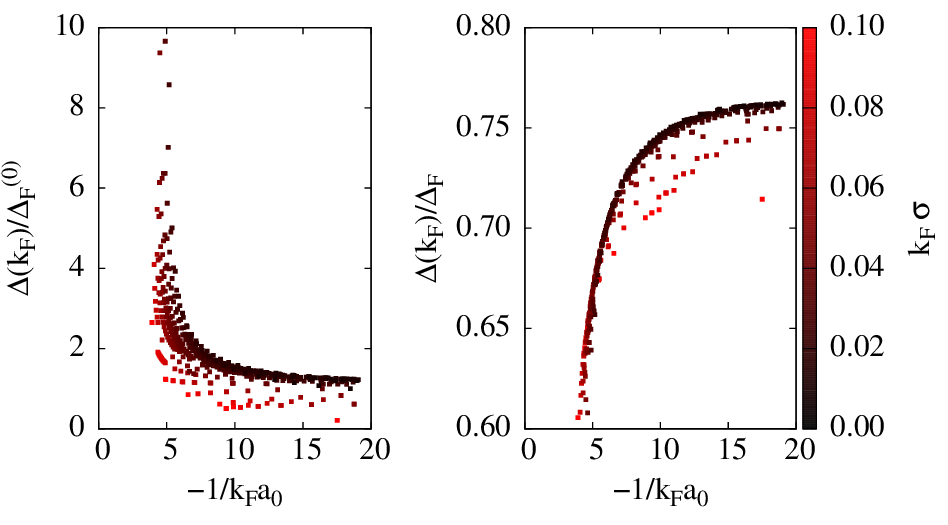}}
\caption{(color online) The superfluid gap $\Delta_{\KF}$ as function
  of $-1/\KF\a0$ obtained for many different values of $\KF$
  ($\KF\sigma\le 0.1$) and $\a0$, for the LJ model.  The color of the
  symbols indicates the $\KF$ values.  Left panel: ratio between the
  gap $\Delta_{\KF}$ and the approximation $\Delta_F^{(0)}$
  (\ref{eq:GapLowdens}).  Right panel: ratio between the gap
  $\Delta_{\KF}$ and the approximation $\Delta_F$
  (\ref{eq:GapApprox}), where a universal dependence on $\KF\a0$ can
  be observed for small $\KF$.  }
\label{fig:gapLJall}
\end{figure}

\begin{figure}
{\includegraphics[width=1.0\columnwidth]{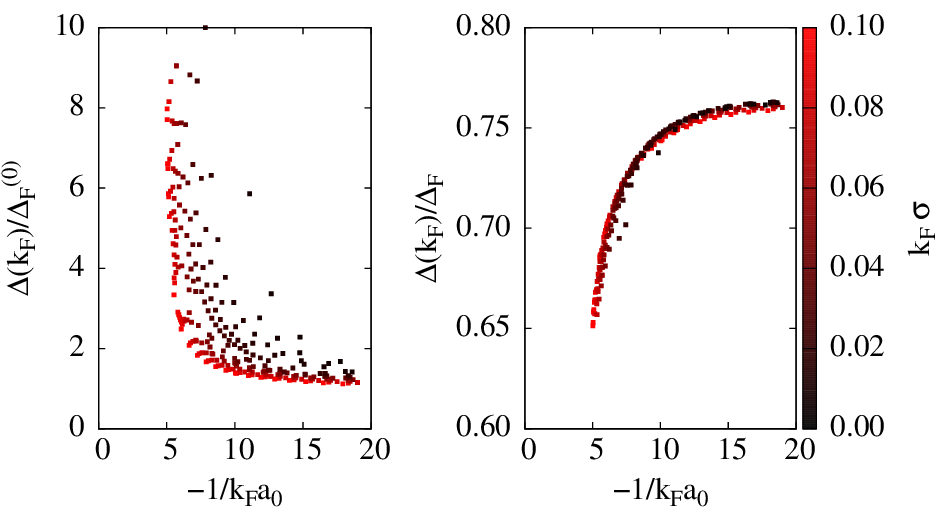}}
\caption{(color online)
Same as Fig.\ref{fig:gapLJall} for the SW model.
}
\label{fig:gapHCall}
\end{figure}

In order to disentangle medium and finite-range corrections, we show
in the right and left panels of Figs.~\ref{fig:gapLJall} (LJ model)
and \ref{fig:gapHCall} (SW model) the ratios $\Delta_{\KF}/\Delta_{F}$
and $\Delta_{\KF}/\Delta^{(0)}_{F}$ as defined in
Eqs. (\ref{eq:GapApprox}) and (\ref{eq:GapLowdens}), respectively,
plotted as function of $-1/\KF\a0$.  The color of the symbols encodes
the $\KF$ value of the corresponding data point, between red for
$\KF\sigma=0.1$ and black for $\KF\to 0$ -- the low density regime
which usually is assumed to be universal in the sense that all
quantities depend on $-1/\KF\a0$ only.  Data for $\KF\sigma>0.1$ are
not shown in these figures.  The two ratios $\Delta_{\KF}/\Delta_{F}$
and $\Delta_{\KF}/\Delta^{(0)}_{F}$ give information on two different
effects: The ratio $\Delta_{\KF}/\Delta_F$ is an assessment of the
accuracy of the approximations leading to Eqs. (\ref{eq:GapApprox})
and reflects the importance of the momentum dependence of the pairing
interaction. Evidently, this effect is visible and depends little on
the density, the ratio does not seem to go to unity with decreasing
density.  This is simply a consequence of our findings of
Figs. \ref{fig:vpairofk}: No matter how small the Fermi momentum is,
the momentum dependence of the pairing interaction is never given by
the naive application (\ref{eq:Wofk}) of low-density arguments but
rather reflects genuine many-body physics. For the range of densities
and coupling strengths, $\Delta_{\KF}/\Delta_F$ is between $60\%$ and
$75\%$, but there is a rapid drop of $\Delta_{\KF}/\Delta_F$ for
$-1/\KF\a0<5$, as the system becomes unstable against dimerization.
In other words, neglecting the momentum-dependence of the pairing
interaction, and thus the finite range of the interaction, leads to an
overestimation of $\Delta_{\KF}$ that becomes more severe as one
approaches the dimerization instability.  We also note that, apart for
some values for large $\KF$, $\Delta_{\KF}/\Delta_F$ collapses on a
single curve, {\em i.e.\/} it is characterized by a universal dependence on
$\KF\a0$.  The curves are identical for the LJ and the SW model.

The difference between our microscopic calculation and the low-density
expression (\ref{eq:GapLowdens}) is much more visible, see the left
panels of Figs. \ref{fig:gapLJall} and \ref{fig:gapHCall}. This effect
has, of course, to do with the divergence of the in-medium scattering
length as discussed above. However, even far from the singularity, we
see very little evidence of an approach to a ``universal'' behavior.
Evidently one must go to much lower densities. We have been able to
carry out our ground state calculations down to densities of
$10^{-11}\,\sigma^{-3}$. At such low densities, the gap itself becomes
many orders of magnitude smaller than the Fermi energy, {\em cf.\/}
Figs. \ref{fig:GapSWLJbare}. Calculations at even lower densities are
therefore only of methodological interest.

\section{Summary}

We have in this paper reported ground state calculations for
low-density Fermi gases described by two model interactions, an
attractive square-well potential and a Lennard-Jones potential. We
have used the optimized Fermi-Hypernetted Chain integral equation
method which has been proved to provide, in the density regimes of
interest here, an accuracy better than one percent \cite{ljium}.  We
have examined the low-density expansion of the energy for a local
correlation operator, written in the conventional Jastrow-Feenberg
form (\ref{eq:wavefunction}). Our formal results also apply for
fixed-node Monte Carlo calculations for wave functions written in that
form. Of course, such a result can only be obtained with optimized
correlation functions. Using a parametrized Jastrow function instead
leads to an unpredictable answer of which one can only say that it
should lie above the expansion (\ref{eq:lowdensfhnc0}). We have
demonstrated that a locally correlated wave function does not
reproduce the exact low-density limit and have cured the problem by
adding the second-order CBF corrections. We have also demonstrated
that already at values of $\a0\KF\approx 10^{-2}$ the third term in
the expansion (\ref{eq:lowdensFermi}) is overshadowed by
model-dependent corrections.

The most interesting result of our work is the appearance of a
divergence of solutions of the FHNC-EL equations well before the
divergence of the vacuum scattering length $\a0$ of the interaction
potential. This makes a statement on both the physics we are
describing and on many-body methodology.

The physics has been described at the end of section
\ref{ssec:energetics}. In the simplest terms it says that the
scattering length $\a0$ appearing in Eq. (\ref{eq:GapApprox}) should
be replaced by the in-medium scattering length $a$ and that the
relationship between $a$ and $\a0$ is non-universal and depends on
many-body effects. Within the medium, the bare interaction is
supplemented by the induced interaction $ w_{\rm I}(r)$, {\em cf.\/}
Eq. (\ref{eq:inducedFermi0}) or (\ref{eq:Wofr}), which describes
phonon exchange. This density-dependent correction to the bare
interaction changes the interaction strength at which a bound state
appears. Thus, the appearance of a divergence of $a$ as a function of
$\a0$ is expected: As a function of coupling strength, the vacuum
scattering length has a singularity somewhat above the strength where
we find the singularity.  The closer the bare interaction is to the
appearance of the bound state, the smaller the correction needs to
be. Since the induced interaction depends on the density, the
singularity appears at lower density for stronger interactions.

We have studied, in the stable regime, the superfluid gap and its
dependence on the density and the interaction strength. Two
corrections apply to low density expansions: medium corrections and
finite-range corrections. We have shown that the most important
finite-range corrections are a direct manifestation of the many-body
nature of the system.  For low densities, the gap can be reasonably
well approximated by neglecting finite-range corrections but
accounting for medium corrections, but the deviations from the full
numerical solution increase on approaching the dimerization
instability.

The second statement is about many-body methodology. The divergence of
solutions of the FHNC-EL equations is in analogy to the spinodal
instability discussed in Sec. 4A, and in many places in the earlier
literature, that the (F)HNC-EL equations for the {\em homogeneous\/}
system have no solutions if $F_0^s < -1$, {\em i.e.\/} if the system
is unstable in the particle-hole channel. ( The fact that this does
not come out exactly (see Eq. (4.4)) is a consequence of the ``fixed
node approximation'' (2.2).)  Here, we prove that the (F)HNC-EL
equations have no solutions if the ground state is unstable against
dimerization, {\em i.e.\/} if the system is unstable in the
particle-particle channel.  This is a unique feature of theories like
FHNC-EL that have the topological completeness of parquet diagrams.

A more appropriate variational wave function for situations with
strong pairing is perhaps an antisymmetrized product of pair functions
as proposed, among others, in Refs. \cite{YangClarkBCS,astraPRL04,%
  bajdichPRL06,PhysRevB.77.115112}. It is well known that the
projection of the BCS state onto a fixed particle number, {\rm i.e.\/}
the states $\ket{\rm BCS^{(N)}}$ can, in coordinate space, be written in
such a form \cite{Schrieffer1999}.  Unfortunately, such a $N$-particle
state does not lead, in the limit of weak coupling, to a BCS-type
theory and a BCS equation (\ref{eq:gap}).  This is easily seen by
considering the stability of the normal state. To make our point, it is
sufficient to consider a weakly interacting system. To guarantee
normalization, we express the Bogoliubov amplitudes in terms of the
real phase angle $\eta_\kvec $: $ u_\kvec \equiv \cos \eta_\kvec,$ and
$ v_\kvec \equiv\ \sin \eta_\kvec$.  Now consider the {\em
  stability\/} of the normal ground state against pairing.  For the
original BCS state one obtains the familiar expression
\cite{BeliaevLesHouches}

\begin{widetext}
\begin{equation}
\left.
\frac{\delta ^2}{\delta\eta_\kvec\delta\eta_{\kvec'}}
\Bra{\rm BCS}H-\mu N\Ket{\rm BCS}
\right|_0\nonumber\\
= (1-2n_0(k))(1-2n_0(k'))\left[2\left|e_k-\mu\right|\delta_{\kvec\kvec'}
+ \bra{\kvec\uparrow,-\kvec\downarrow}
V \ket{\kvec'\uparrow,-\kvec'\downarrow}\right]\,,
\end{equation}
where $n_0(k)=\theta(\KF-k)$ is the normal Fermi distribution.

If we use, instead, the number-projected state
$\Ket{{\rm BCS}^{(N)}}$ we get
\begin{equation}
\left.
\frac{\delta ^2}{\delta\eta_\kvec\delta\eta_{\kvec'}}
\frac{\Bra{{\rm BCS}^{(N)}}H-\mu N\Ket{{\rm BCS}^{(N)}}}
{\left\langle{\rm BCS}^{(N)}\mid {\rm BCS}^{(N)}\right\rangle}
\right|_0
=- \bra{\kvec\uparrow,-\kvec\downarrow}
V \ket{\kvec'\uparrow,-\kvec'\downarrow}
\end{equation}
for $k>\KF$ and $k'<\KF$ or vice versa, and zero otherwise.
See Ref. \onlinecite{HNCBCS}, Eqs. (2.7), (2.8).
\end{widetext}
Thus, the number-projected BCS state, while useful for strongly
coupled pairs, {\em i.e.\/} typically beyond the first singularity of
the (in-medium) scattering length, does not reproduce the correct
stability condition in the weakly-coupled limit. Full optimization of
such a wave function containing both Jastrow correlations and an
antisymmetrized product of pair wave functions requires first to
derive the diagrammatic expansion, which goes beyond the scope of this
paper.

\appendix

\section{Low-density expansions for fermions}
\subsection{Low density limit for local correlations}
\label{app:lowdens}

In the limit of low densities, the exact energy per particle is given
by the expansion (\ref{eq:lowdensFermi})
\cite{HuangYang57,Landau5}. The wave function containing a local
correlation operator (\ref{eq:wavefunction}) is not exact, it is of
interest to determine the consequences of this approximation for the
low-density expansion.

There are two ways to derive the low--density expansion for the wave
function (\ref{eq:wavefunction}): One is to assume that the
interaction is weak and has a Fourier transform. Then one can proceed
in the same way as in Ref.  \onlinecite{parquet1}, {\em i.e.\/}

\begin{figure}[h]
\centerline{\includegraphics[width=0.8\columnwidth]{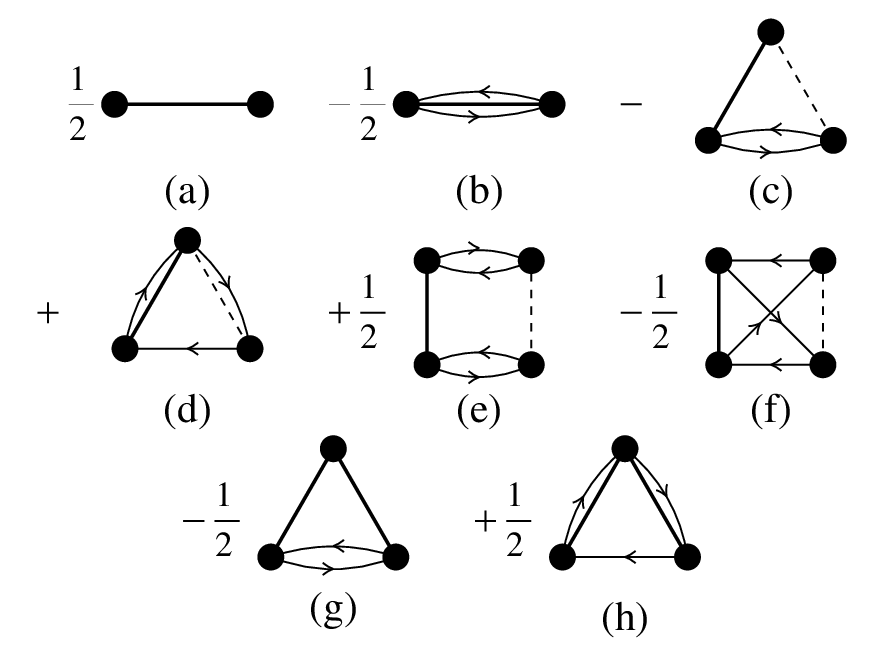}}
\caption{The energy functional to second order. The usual conventions
  concerning diagrams apply. The heavy solid line represents a
  Clark-Westhaus two-body interaction, $v_{\rm CW}(r_{ij})$, and the
  pair of heavy lines in diagrams (g) and (h) the three-body term
  $\hm4\nabla_i\Gamma_{\rm dd}(r_{ij})\cdot\nabla_i\Gamma_{\rm
    dd}(r_{ik})$, see Eq. (\ref{eq:TCW}).}
\label{fig:fig2}
\end{figure}

\begin{itemize}
\item Derive the cluster-expansion of the energy keeping all terms
  that are (a) of second order in the correlation function $f^2(r) -1
  \equiv e^{u_2(r)}$, or (b) of first order in the potential and first
  order in the correlation function
\item Derive the Euler equation and express the correlation function to
  first order in the potential
\item Re-insert the optimal correlation function in the energy
\item Collect all terms that can be written as powers $\KF\a0 $.
\end{itemize}

If we also permit hard-core interactions, we must formulate the
expansion in terms of the dynamical correlation function $\Gamma_{\rm
  dd}(r)$ and take into account that the solution of the Euler
equation is density dependent.  The analytic expression for the
relevant energy contributions is given in Eqs. (\ref{eq:EJF}),
(\ref{eq:EQ}) and (\ref{eq:TCW}).  The only additional simplification
is that we can set, to leading order in $\Gamma_{\rm dd}(r)$, the
factor $S^2_{\rm F}(q)/S(q)-1 \approx S_{\rm F}(q)-1$ in $e_{\rm
  Q}^{(1)}$.  The resulting diagrammatic representation of the energy
expansion is shown in Fig. \ref{fig:fig2}.

In the low-density limit, $v_{\rm CW}(r)$ is much shorter ranged than
both the exchange function $\ell(r\KF)$ and the dynamic correlation
function $\Gamma_{\rm dd}(r)$. $\tilde v_{\rm CW}(k)$ can therefore be
replaced by its long--wavelength limit $\tilde v_{\rm CW}(0+)$ which is,
in turn, related to the scattering length, see Eq. (\ref{eq:vcw0}).
Then, diagrams (b), (d), and (f) are $1/\nu$ times diagrams (a), (c),
and (e), respectively. For the correlation function
(\ref{eq:EulerLow0}), all diagrams shown in Fig. \ref{fig:fig2} can be
calculated exactly in terms of the long-wavelength limit $\tilde
v_{\rm CW}(0+)$. The calculation is somewhat tedious but
straightforward and leads to the aforementioned result
(\ref{eq:lowdensfhnc0}). However, this is {\em not\/} the exact
solution because the exchange diagram (f) has not been treated
exactly.

To derive the exact low-density limit, we start from the
diagrammatic expansion up to second order is shown in Fig.
\ref{fig:fig2}. Carrying out the variation
with respect to $\sqrt{1+\Gamma_{\rm dd}(r)}$ we get
\begin{widetext}
\begin{eqnarray}
&&\hm{}
\nabla\cdot\left[g_F(r)\nabla\sqrt{1+\Gamma_{\rm dd}(r)} \right] + \hm2
\left[\nabla^2\Gamma_{\rm dd}(k)(\SF(k)-1)\right]^{\cal F}(r)
\sqrt{1+\Gamma_{\rm dd}(r)}
\nonumber\\
&&\qquad+ \frac{\hbar^2}{2m\nu^2}\biggl[\nabla\cdot
\int d^3r'
 \ell(r'\KF )\ell(r\KF )
\ell(|\rvec'-\rvec|\KF )\nabla_{\rvec}\Gamma_{\rm dd}(|\rvec'-\rvec|)\biggr]
\sqrt{1+\Gamma_{\rm dd}(r')} \nonumber\\
&&= \left[v(r) + \left[\tilde v_{\rm CW}(k)(S^2_F(k)-1)\right]^{\cal F}(r) +
\Delta X_{\rm ee}'(r)\right]\sqrt{1+\Gamma_{\rm dd}(r)}\label{eq:temp0}
\end{eqnarray}
where $(\Delta X'_{\rm ee})_1(r)$ is represented by the two diagrams
shown in Fig. \ref{fig:Xee}, with the dashed line replaced by $v_{\rm
  CW}(r)$. In leading order in $\KF $, we can replace $g_F(r) = g_F(0)
=1/2$ and expand
\begin{eqnarray}
(\Delta X'_{\rm ee})_1^{(3)}(r)&=&\frac{2\rho}{\nu^2}\int d^3r_3v_{\rm CW}(r_{13})
\ell(r_{12}\KF )\ell(r_{13}\KF )\ell(r_{23}\KF )
=\frac{2\rho}{\nu^2}\int d^3r_3v_{\rm CW}(r_{13})\ell^2(r_{23}\KF )+
\rho\times{\cal O}(\KF ^2)\nonumber\\
&=& -\frac{2}{\nu}
\left[\tilde v_{\rm CW}(k)(1-\SF(k))\right]^{\cal F}(r) +
\rho\times{\cal O}(\KF ^2)
\end{eqnarray}
\end{widetext}
and similarly
\begin{equation}
(\delta X'_{\rm ee})_1^{(4)}(r)=  \frac{1}{\nu}
\left[\tilde v_{\rm CW}(k)(1-\SF(k))^2\right]^{\cal F}(r) +
\rho\times{\cal O}(\KF ^2)\,.
\end{equation}
Finally, we combine the last two terms in Eq. (\ref{eq:temp0}) to
$g_F(0)\left[\tilde v_{\rm CW}(k)(1-\SF(k))\right]^{\cal F}(r)$.

\begin{figure}
\centerline{\includegraphics[width=1.0\columnwidth]{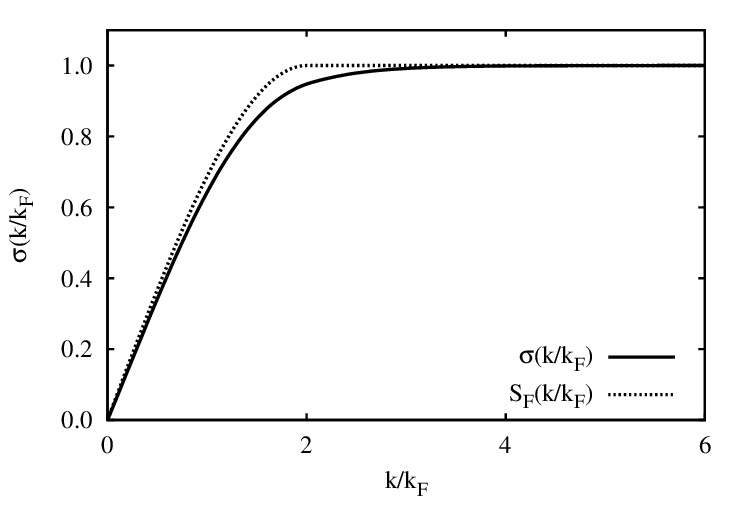}}
\caption{The universal function $\sigma(k/\KF)$, defined by
$\tilde\Gamma_{\rm dd}(k) = \displaystyle
-\frac{\tilde v_{\rm CW}(k)\sigma(k/\KF)}{t(k)}$
(see text) is shown (full line), in comparison to $\SF(k)$ (dotted line).}
\label{fig:sigma}
\end{figure}

Multiplying the equation with $\sqrt{1+\Gamma_{\rm dd}(r)}$, using
\begin{align*}
&\sqrt{1+\Gamma_{\rm dd}(r)}\nabla^2 \sqrt{1+\Gamma_{\rm dd}(r)}\\
 = &\half\nabla^2\Gamma_{\rm dd}(r)
 - \left|\nabla \sqrt{1+\Gamma_{\rm dd}(r)}\right|^2\,,
\end{align*}
and keeping only first order terms in $\Gamma_{\rm dd}(r)$
reduces Eq. (\ref{eq:temp0}) to
\begin{widetext}
\begin{align}
&\frac{\hbar^2}{2m}g_F(0)\nabla^2 \Gamma_{\rm dd}(r)
 + \frac{\hbar^2}{2m}\nabla^2\left[\tilde \Gamma_{\rm dd}(k)(\SF(k)-1)
\right]^{\cal F}\!\!(r)
+ \frac{\hbar^2\rho}{2m\nu^2}\nabla_{\rvec}\cdot
\int d^3r'\left[\ell(rk_F)\ell(r'k_F)\ell(|\rvec-\rvec'|k_F)
\nabla_{\rvec}  \Gamma_{\rm dd}(|\rvec-\rvec'|)\right]\nonumber\\
&= g_F(0)\left[\tilde v_{\rm CW}(k)\SF^2(k)\right]^{\cal F}(r)\,.
\label{eq:temp1}
\end{align}
\end{widetext}
The approximation leading to Eq. (\ref{eq:EulerLow0}) is to replace the
third term by
\[ \ell(rk_F)\ell(r'k_F)\ell(|\rvec-\rvec'|k_F)\approx
\ell^2(r'k_F)\,.\]
This is legitimate if the range of $\Gamma_{\rm dd}(r)$ is small
compared to that of $\ell(rk_F)$ which is not necessarily true.
To assess the importance of this term, we write
\begin{equation}
\tilde\Gamma_{\rm dd}(k) = -\frac{\tilde v_{\rm CW}(k)\sigma(k/\KF)}{t(k)}
\label{eq:optmany}
\end{equation}
with $\sigma(k)\propto k$ as $k\rightarrow 0$ and
$\sigma(k)\rightarrow 1$ as $k\rightarrow \infty$. In the long
wavelength limit $v_{\rm CW}(k)\approx \tilde v_{\rm CW}(0+)$, the
function $\sigma(k/\KF)$ becomes universal. We have solved the Euler
equation (\ref{eq:temp1}) in that limit, the resulting function
$\sigma(k/\KF)$ is shown in Fig. \ref{fig:sigma}. The deviation from
$\SF(k)$ is small but visible. Calculating the energy correction to
second order in $(a\KF)$ leads to a coefficient $1.5519$ instead of
$1.5415$. The smallness of the effect is plausible because diagram (g)
contributes only a few percent to the total energy correction.

\subsection{Density Dependence of the Correlation  Functions}

We have in Eq. (\ref{eq:optmany}) derived the connection between the
dynamical correlation function $\Gamma_{\rm dd}(r)$ and the effective
interaction $v_{\rm CW}(r)$. To complete the calculation of the
low-density limit of the energy for local correlation operators, we
must establish the connection between $\tilde v_{\rm CW}(0+)$ and the
scattering length $a_0$ because $\tilde v_{\rm CW}(0+)$ is calculated
with the optimal correlation function (\ref{eq:optmany}) of the
many-body problem at {\rm finite density,} and not with the solution
of the vacuum scattering equation (\ref{eq:scatteq}).  To calculate
the correction, let
\begin{align*}
\delta\psi(r) &= \psi(r)-\sqrt{1+\Gamma_{\rm dd}(r)} \\
\delta\tilde \psi(k) &= \frac{\tilde v_{\rm CW}(k)(\SF(k)-1)}{2t(k)}\,.
\end{align*}
Then
\begin{eqnarray}
\tilde v_{\rm CW}(0) &=& \rho\int d^3r\left[\frac{\hbar^2}{m}
\left|\nabla \psi(r)\right|^2
+v(r)\psi^2(r)\right]\nonumber\\
&+& 2 \rho\int d^3r\left[\frac{\hbar^2}{m}
\nabla \psi(r)\cdot\nabla\delta\psi(r)
+v(r)\psi(r)\delta\psi(r)\right]\nonumber\\
&+& \rho\int d^3r\left[\frac{\hbar^2}{m}
\left|\nabla \delta \psi(r)\right|^2
+v(r)\delta\psi^2(r)\right]\nonumber\\
 &=&\frac{4\pi\rho\hbar^2}{m}a_0 +  \rho\int d^3r\frac{\hbar^2}{m}
\left|\nabla \delta \psi(r)\right|^2 + {\cal O}(v^3)\,.
\end{eqnarray}
The last line follows because $v(r)|\delta\psi(r)|^2$ is of order
${\cal O}(v^3)$, and the mixed term is zero due to the scattering equation
(\ref{eq:scatteq}). Since we have eliminated the potentially
singular bare potential, we can expand the kinetic energy term to get
\begin{equation}
\tilde v_{\rm CW}(0+) = \frac{4\pi\rho\hbar^2}{m}a_0
+ \frac{1}{2} \int\frac{d^3 k}{(2\pi)^3\rho}\frac{\tilde v_{\rm CW}^2(k)}{t(k)}
\left[\SF(k) -1\right]^2\,.
\label{eq:vcw0app}
\end{equation}
Collecting all results, one finds
\begin{equation}
\frac{E}{N} =\frac{\hbar^2 \KF ^2}{2m}\left[\frac{3}{5}
+ \frac{2}{3}\frac{ a \KF }{\pi} +  1.53 517
\left(\frac{a \KF }{\pi}\right)^2 +\ldots\right]\,.
\label{eq:appfhnc0}
\end{equation}
The numerical factor $1.53517$ is to be compared with the factor
$4(11-2\ln 2)/35 = 1.098$ of Eq. (\ref{eq:lowdensFermi}).  We
emphasize again that the result (\ref{eq:lowdensfhnc0}) also applies
for the ``fixed-node'' approximation because the terms, where that
approximation deviates from our expansion, are of at least fourth
order in the potential strength.

\subsection{Correlated Basis Functions Corrections}

Once we have determined that a local correlation operator of the form
(\ref{eq:wavefunction}) does not lead to the correct low-density limit
(\ref{eq:lowdensFermi}) of the correlation energy, the question arises
what corrections to that wave function are necessary. We show here
that second-order CBF perturbation theory (\ref{eq:E2CBF}) is
sufficient to retrieve the exact low-density expansion.  In the
low-density (local) approximation $\tilde {\cal W}(q)$ and $\tilde
{\cal N}(q)$ are given by Eqs. (\ref{eq:NWloc}).

There are normally significant numerical cancellations between the two
terms in the numerator of Eq. (\ref{eq:E2CBF}).  However, for the
purpose of discussing formal properties, we may expand the
numerator. The term containing two matrix-elements $\left\langle
  pp'\right|{\cal W}\left|hh'\right\rangle_a$ is just the ordinary
expression of second-order perturbation theory \cite{Landau5}. None of
the other terms contains an energy denominator, they can therefore be
written in terms of the diagrams of the JF method.

We carry out the calculation for just the direct term (notice that the
double-exchange is equal to the direct term):
\begin{widetext}
\begin{eqnarray}
\delta E_2^{\rm dir}
&=& -\frac{1}{2}\int \frac{d^3q}{(2\pi)^3\rho}
\int\frac{ d^3h d^3h'}{V_F^2}
n(h) n(h')\bar n(h+q)\bar n(h'+q)
\frac{\left|\tilde {\cal W}(q)
+ \frac{1}{2}\left[t_p + t_{p'}-t_h -t_{h'}\right]\tilde {\cal N}(q)
\right|^2}
{t_p + t_{p'}-t_h -t_{h'}}\nonumber\\
&=&
 -\frac{1}{4}\int \frac{d^3q}{(2\pi)^3\rho}
\left[\frac{\tilde v_{\rm CW}^2(q)}{t(q)}I(q)
+ 2\tilde v_{\rm CW}(q)\tilde\Gamma_{\rm dd}(q)\SF^2(q)
+ t(q)\tilde\Gamma_{\rm dd}^2(q)\SF(q)\right]\label{eq:temp2}
\end{eqnarray}
\end{widetext}
where $V_F = \frac{4\pi}{3}\KF^3$, $\bar n(k) = 1-n(k)$,
and
\begin{equation}
I(q) = 2 t_q \int\frac{ d^3h d^3h'}{V_F^2}
\frac{ n(h) n(h')\bar n(h+q)\bar n(h'+q)}
{t_p + t_{p'}-t_h -t_{h'}}\,.
\end{equation}
Expanding the second term as
\begin{align}
-\frac{1}{2}\int \frac{d^3q}{(2\pi)^3\rho}
\tilde v_{\rm CW}(q)\tilde\Gamma_{\rm dd}(q)
\big[&(\SF(q)-1)^2(q)\label{eq:temp3}\\
&+2(\SF(q)-1)+1\big],\nonumber
\end{align}
we identify in the first term the negative diagram (e) and in the second
term the negative diagram (c) of Fig. \ref{fig:fig2}.
Likewise, expanding the last term of Eq. (\ref{eq:temp2})
as
\begin{equation}
-\frac{1}{4}\int \frac{d^3q}{(2\pi)^3\rho}
t(q)\tilde\Gamma_{\rm dd}^2(q)\left[(\SF(q)-1)+1\right],
\label{eq:temp4}
\end{equation}
we identify in the first term the negative diagram (g) of Fig. \ref{fig:fig2}.
Thus, these terms cancel exactly all direct, two-line diagrams shown in Fig.
\ref{fig:fig2}.

The last terms in Eqs. (\ref{eq:temp3}) and (\ref{eq:temp4}) cannot
be represented by legitimate diagrams of JF-theory, they are combined to
\begin{align}
&-\frac{1}{4}\int \frac{d^3q}{(2\pi)^3\rho}
\left[t(q)\tilde\Gamma_{\rm dd}^2(q)+2\tilde v_{\rm CW}(q)\Gamma_{\rm dd}(q)
\right]\nonumber\\
=& -\frac{1}{4}\int \frac{d^3q}{(2\pi)^3\rho}
\left[\frac{\tilde v_{\rm CW}^2(q)}{t(q)}
\left(\SF(q)-1\right)^2 + \frac{\tilde v_{\rm CW}^2(q)}{t(q)}\right]\,.
\label{eq:temp5}
\end{align}
The first term is seen to cancel the correction
(\ref{eq:vcw0app}). The second term is combined with the first term in
Eq. (\ref{eq:temp2}) to give the textbook result \cite{Landau5}
\[
 -\frac{1}{4}\int \frac{d^3q}{(2\pi)^3\rho}
\frac{\tilde v_{\rm CW}^2(q)}{t(q)}\left[I(q)-1\right]\,.
\]
The treatment of the exchange terms is done along the same
lines and will not be repeated here.

\section{Numerical solution of the gap equation}
\subsection{General strategy}

We present in this appendix details of our solution method for the gap
equation which is numerically delicate.

The gap equation
\begin{align}
\Delta(k) &= -\frac{1}{2}\sum_{k'}{\cal P}_{\kvec,\kvec'}\frac{\Delta(k')}{
\sqrt{(\epsilon_{k'}-\mu)^2 + \Delta^2(k')}}\nonumber\\
 &= -\frac{4\pi}{2\,(2\pi)^3\rho}\int dk' {k'}^2 P(k,k')\frac{\Delta(k')}{
\sqrt{(\epsilon_{k'}-\mu)^2 + \Delta^2(k')}}\,
\label{eq:gapeq}
\end{align}
is highly non-linear and a simple iteration procedure does not
converge. For values $\Delta(k)\ll \EF$, the integrand is narrowly
peaked at the Fermi momentum and an adaptive mesh is necessary to
reach a reliable numerical accuracy.  Above, we have introduced
the angular average
\begin{equation}
P(k,k') = N\int\frac{ d^2\Omega_{\kvec,\kvec'}}{4\pi}{\cal P}_{\kvec,\kvec'}\,.
\end{equation}

A very rapidly converging algorithm is as follows:
Consider the generalized eigenvalue problem
\begin{align}
&\lambda(\xi)\Delta_{n+1}(k,\xi) =\\
& -\frac{1}{4\pi^2\rho}\int dk' {k'}^2 P(k,k')
\frac{\Delta_{n+1}(k')}{\sqrt{(\epsilon_{k'}-\mu)^2 + \xi^2\Delta_n^2(k')}}. \nonumber
\label{eq:eigengap}
\end{align}
 To shorten the
equations, let
\begin{equation}
E_n(k,\xi) \equiv\sqrt{(\epsilon_{k}-\mu)^2 + \xi^2\Delta_n^2(k)}\,.
\end{equation}
The problem can be mapped to a regular symmetric eigenvalue problem
by defining
\begin{equation}
D_{n+1}(k,\xi) =  \frac{\Delta_{n+1}(k')}{\sqrt{E_n(k,\xi)}}.
\end{equation}
Then, the eigenvalue problem is
\begin{align*}
&\lambda(\xi) D_{n+1}(k,\xi) = \\
&-\frac{1}{4\pi^2\rho}\int dk' {k'}^2
\frac{P(k,k')}{\sqrt{E_n(k,\xi) E_n(k',\xi)}}
D_{n+1}(k',\xi).
\end{align*}

The algorithm is
\begin{enumerate}[(i)]
\item{} Start with a reasonable estimate $\Delta_0(k')$, {\em e.g.}
a constant
\item{} Solve the above eigenvalue problem as a function of the
  scaling parameter $\xi$ and find the value $\xi_1$ for which the
  lowest eigenvalue is 1. Along with a trial solution, we can
  calculate the derivative with respect to $\xi$, see below.
\item{} Scale the corresponding eigenfunction such that
$\Delta_{n+1}(\KF) = \xi\Delta_n(\KF)$.
\item{} Go to step (ii) and repeat until convergence which is obtained
for $\xi = 1$.
\end{enumerate}

\subsection{Adaptive mesh}

To deal with the strongly peaked denominator
$E(k,\xi)$, we introduce a transformation $k = k(x)$
that generates a dense set of points around the Fermi momentum.
Then
\begin{equation}
\Delta(k) =  -\frac{1}{4\pi^2\rho}
\int {k'}^2 P(k,k')\frac{\Delta(k')}{
\sqrt{(\epsilon_{k'}-\mu)^2 + \Delta^2(k')}}
\frac{dk'}{dx} dx\,.\label{eq:gap0}
\end{equation}
Now find a function $k(x)$ such that
\begin{equation}
\frac{dk}{dx}\frac{1}{\sqrt{(\epsilon_{k}-\mu)^2 + \Delta^2(k)}}
\label{eq:erat}
\end{equation}
is almost constant. The function of choice is
\begin{eqnarray}
x(k) &=& \frac{\KF^2}{\EF}\int_0^{\kappa} \frac{\kappa' d\kappa'}
{\sqrt{({\kappa'}^2-1)^2+\delta^2}}\\
&=&\frac{\KF^2}{2\EF}
\log\left(\frac{\sqrt{(\kappa^2-1)^2+\delta^2}+ \kappa^2-1}
{\sqrt{1+\delta^2}-1}\right)\,,
\end{eqnarray}
where $\delta = \Delta(k_F)/\EF$ and $\kappa = k/\KF$. This
choice has the property
\begin{equation}
\frac{dx}{dk} = \frac{k}{\EF\sqrt{(k^2/\KF^2-1)^2+\delta^2}},
\end{equation}
{\em i.e.\/}
\[\sqrt{(\epsilon_{k'}-\mu)^2 + \Delta^2(k')}
\frac{dk'}{dx}\]
is almost constant.

We can also obtain a closed-form expression for $k(x)$:
\begin{equation}
\left(\frac{k}{\KF}\right)^2 = 1 +
\sqrt{1+\delta^2}\sinh(2 \EF x/\KF^2) - \cosh(2 \EF x/\KF^2)\,.
\end{equation}

\bigskip

\begin{acknowledgments}

  This work was supported, in part, by the College of Arts and
  Sciences, University at Buffalo SUNY, and the Austrian Science Fund
  Projects P21264 and I602 (to EK) and P23535 (to REZ).  Additional
  support was provided by the Spanish Ministry of Science and
  Education (MEC) through project FIS2011-28617-C02-01 and from the
  Qatar National Research Fund \# NPRP 5 - 674 - 1 - 114. Discussions
  with J. Boronat are also acknowledged.

\end{acknowledgments}
\bibliography {papers}
\bibliographystyle{apsrev4-1}

\end{document}